\documentclass[12pt]{article}
\usepackage{setspace}
\usepackage{amsmath, amssymb, amsthm, bbm}
\usepackage[usenames,dvipsnames]{xcolor}
\definecolor{dukeblue}{rgb}{0.0, 0.0, 0.61}
\usepackage[round]{natbib}
\usepackage[hyperfootnotes=false]{hyperref}
\hypersetup{
	colorlinks,
	citecolor=dukeblue,
	linkcolor=dukeblue,
	urlcolor=dukeblue}
\usepackage[a4paper]{geometry}
\usepackage[utf8]{inputenc}
\usepackage[flushleft]{threeparttable}
\usepackage{indentfirst,graphicx,comment,amsmath,amsfonts,bbm,standalone,float,array,multirow,enumitem,mathtools,subcaption,amssymb,bm,times,chngcntr,amsthm,longtable,threeparttablex,dcolumn,booktabs,lscape,caption,rotating,parskip,comment,pdflscape}
\usepackage{xurl}
\usepackage{tgpagella}
\usepackage[font={small,bf}]{caption}
\newtheorem {theorem}{Theorem}

\newtheorem {proposition}{Proposition}

\newtheorem {assumption}{Assumption}
	\newcommand\blfootnote[1]{%
	\begingroup
	\renewcommand\thefootnote{}\footnote{#1}%
	\addtocounter{footnote}{-1}%
	\endgroup
}

\usepackage{tikz}
\newcommand*\circled[1]{\tikz[baseline=(char.base)]{
		\node[shape=circle,draw,inner sep=2pt] (char) {#1};}}

\begin{document}
\title{Difference-in-Differences with a Misclassified Treatment}
\author{Akanksha Negi$^\dag$ and Digvijay S. Negi$^\ddag$}
\date{This Version: August, 2022 \\ {\small First Version: December, 2021}}
\maketitle

\blfootnote{$^\dag$Department of Econometrics and Business Statistics, Monash University. Email: \href{mailto:akanksha.negi@monash.edu}{akanksha.negi@monash.edu}. $^\ddag$Indira Gandhi Institute of Development Research, Mumbai. Email: \href{mailto:digvijay@igidr.ac.in}{digvijay@igidr.ac.in}. \\
	$^\ast$We would like to thank seminar participants at EWMES 2021 and ESAM 2022 for comments and suggestions on earlier versions of this paper.}

\begin{spacing}{1.0}
	\begin{abstract}
		This paper studies identification and estimation of the  average treatment effect on the treated (ATT) in difference-in-difference (DID) designs when the variable that classifies individuals into treatment and control groups (treatment status, $D$) is endogenously misclassified. We show that misclassification in $D$ hampers consistent estimation of ATT because 1) it restricts us from identifying the truly treated from those misclassified as being treated and 2) differential misclassification in counterfactual trends may result in parallel trends being violated with $D$ even when they hold with the true but unobserved $D^\ast$. We propose a solution to correct for endogenous one-sided misclassification in the context of a parametric DID regression which allows for considerable heterogeneity in treatment effects and establish its asymptotic properties in panel and repeated cross section settings. Furthermore, we illustrate the method by using it to estimate the insurance impact of a large-scale in-kind food transfer program in India which is known to suffer from large targeting errors.  
	\end{abstract}
	
	
	
	\newpage 
\end{spacing}

\section{Introduction}\label{intro}

The degree of measurement error in economic data and its influence on parameter estimates have been of much interest to econometricians. One case in particular is measurement error in a binary variable, also known as misclassification. Unlike classical measurement error, misclassification is necessarily non-classical since the error is negatively correlated with the truth \citep{aigner1973regression,bound2001measurement}. This is an important case in the program evaluation literature, of which Difference-in-Differences is a workhorse empirical strategy, because the key regressor of interest, the \textit{treatment}, is often binary in nature.\footnote{Evidence from the literature suggest that ignoring even a small amount of misclassification can have major repercussions for the estimated treatment effects \citep{millimet2011elephant,kreider2010regression}.}

In this paper, we discuss identification and estimation of the average treatment effect on the treated (ATT) in a DID setting when the binary variable that classifies individuals into treatment or control groups (treatment status, $D$) is misclassified. The recent surge in the econometric literature on DID has seen a renewed interest in the identification, estimation, and interpretation of effects using the most common regression specifications. Despite such attention, the effects of a misclassified treatment on the DID estimand and its causal interpretation remain unknown. 

Participation in programs or interventions is particularly prone to mismeasurment due to self-reporting or mistargeting  \citep{bruckmeier2021misreporting,martinelli2009deception,coady2004targeting}. For instance, \citet{meyer2015household} document how misreporting of program receipt and conditional transfers from the government appear to be the biggest threat to household survey data quality for policy evaluation. Given that DID is often applied to data from nationally representative surveys \citep{groen2008effect,buchmueller2011effect, botosaru2018difference}, it becomes all the more important to study the effects of misclassification on the DID estimand and its causal interpretation as the ATT.  

We begin by showing that with $D$, the DID estimand can be decomposed into a sum of ATT for those misclassified as being treated (i.e. $D=1$) and the difference in counterfactual trends between those misclassified as being treated vis-$\grave{a}$-vis the misclassified controls. Therefore, the bias in DID is two-fold and confounds identification of the true ATT because: 1) it restricts us from identifying the truly treated from those misclassified as being treated (unless misclassification probabilities are known) and 2) the counterfactual trends in $D$ may not be parallel between those observed to be treated or non-treated even if parallel trends hold with the true unobserved $D^\ast$. 

The first component of this bias is not new and has appeared in the misclassification literature including \cite{aigner1973regression}, \citet{lewbel2007estimation}, \citet{battistin2011misclassified} among others. The second component, however, is a unique aspect of the DID design where parallel trends play an instrumental role in identifying the ATT. With $D$, parallel trends may not hold under the general case of differential misclassification in the counterfactual trends of the treated and control groups. But if we assume non-differential misclassification in the trends, then proposition \ref{prop1} in our paper presents a key relationship between counterfactual trends in $D$ vis-$\grave{a}$-vis $D^\ast$. It show that when parallel trends hold with $D^\ast$, then they will also hold with $D$ and vice versa except in the odd case where the  probability of misclassification is the same as that of  correct classification. 

We outline our main approach in the context of a flexible linear specification of the potential outcome means and use it to arrive at parametric DID regressions for 2-group 2-period (2$\times$2) panel and repeated cross section settings. Our approach follows from \citet{nguimkeu2019estimation} (NDT thereafter) who study endogenous participation and endogenous misreporting of social programs. We extend NDT to a 2$\times$2 setting and focus on the one-sided misclassification case which only explicitly accounts for errors of exclusion. Our regressions allow for heterogeneity by having interactions between covariates, treatment, and time. We then propose a two-step estimator which corrects for endogenous one-sided misclassification using a partial observability probit model (POP). Identification of the POP parameters and in turn the true ATT is achieved using a single exclusion restriction that only affects the misclassification probability but does not influence the true treatment or the outcome model. While NDT require two exclusion restrictions (one for endogenous treatment and the other for endogenous misclassification), we require only one since the DID design accounts for  endogenous participation as long as it is based on time-invariant unobservables. 

As an illustration of this method, we apply it to study the welfare impact of the Public Distribution System (PDS) which is a large-scale program for targeted distribution of highly subsidized food grains in India. One key argument in favor of in-kind transfers is that they provide implicit insurance to eligible households from commodity price risk \citep{gadenne2021kind,negi2022global}. Using India as the setting, we quantify the welfare benefits of PDS for the poor who are most vulnerable to high food prices. The key challenge in estimating the welfare impact of PDS is that it is known to have large targeting errors where errors of wrong exclusion are disproportionately larger than errors of wrong inclusion \citep{dutta2001targeting,swaminathan2001errors, hirway2003identification,khera2008access,jha2013safety,pingali2019reimagining}. We use the two-step estimator to correct for one-sided mistargeting of the treatment due to errors of wrong exclusion. Although our theory is in terms of misclassification, we show that it can be reformulated to fit the context of mistargeting, demonstrating the versatility of the method in different empirical settings.

Our paper makes three important contributions. The first contribution is to study and characterize the effects of misclassification on the DID estimand and its implications for identifying and estimating the ATT. To the best of our knowledge, this is the first paper to discuss the problem of misclassification within a DID framework and highlight the two important sources of bias in the DID estimand on account of using $D$ rather than $D^\ast$.  We also argue how parallel trends in $D^\ast$ or even $D$ is no longer sufficient for identifying the true ATT and present how they relate to each other under the assumption of non-differential misclassification in counterfactual trends. Our second contribution is to propose a solution for the case of endogenous (or differential) one-sided misclassification in panel and repeated cross-section regressions and to characterize the inconsistency (asymptotic bias) in the first-differenced (FD) and pooled OLS (POLS) estimators of ATT for the panel and repeated cross section regressions, respectively. In doing so, we extend NDT's approach to the DID setting and establish consistency and asymptotic normality of the two-step DID estimators. A salient difference between NDT and us is that we require only one exclusion restriction which affects the misclassification probability but can still allow for an endogenous treatment as long as it is due to time-invariant factors. Our third contribution is empirical in nature where we apply the proposed method to estimate the insurance impact of PDS in the presence of one-sided mistargeting which is a reality in social welfare programs. We find that if the  PDS subsidies were correctly targeted, the welfare impact of the program would have been much larger than what was actually observed.

Our paper contributes to two different strands of the causal inference literature. The most obvious contribution is to the misclassified binary regressor literature with antecedents such as \citet{aigner1973regression}, \citet{mahajan2006identification}, \citet{lewbel2007estimation}, \citet{frazis2003estimating},  \citet{meyer2017misclassification}, \citet{haider2020correcting},  \citet{calvietal2021}, \citet{bollinger1996bounding}, \citet{battistin2011misclassified}, and \citet{ditraglia2019identifying}. More recent contributions include \citet{ura2018heterogeneous}, \citet{yanagi2019inference}, \citet{tommasi2020bounding}, and \citet{acerenza2021marginal}. The other strand concerns the broader DID literature which includes \citet{abadie2005semiparametric}, \citet{botosaru2018difference}, \citet{goodman2018difference}, \citet{de2020two},  \citet{sant2020doubly}, \citet{sun2020estimating}, \citet{callaway2020difference},  \citet{rambachan2022more}, and \citet{wooldridgetwfe} to name a few. 

In particular, our paper is closely related to \citet{botosaru2018difference} in the sense that they also depart from assuming perfect knowledge of the treatment status in DID studies and are rather interested in identifying the ATT when the treatment status in missing in one of the two periods. This is different than the main premise of our paper which assumes the treatment status to be observed with error or misclassified. Our paper is more general than \citet{botosaru2018difference} since the problem of missing treatment status can be accommodated within our framework.\footnote{Suppose the missing treatment status is predicted in one of the two periods (as in \citet{botosaru2018difference}), then the predicted status can be considered to be the observed but misclassified treatment status.} 

The rest of this paper is organized as follows. Section \ref{lit} presents a brief review of the binary misclassified regressor and the DID literatures. Section \ref{did_simple} uses the standard DID framework to characterize the bias in the DID estimand when one uses the misclassified proxy $D$ in place of the true but unobserved treatment, $D^\ast$. This section also shows that under the simplifying assumption of non-differential misclassification, a relationship between parallel trends with $D$ and $D^\ast$ is easily established. Despite that, the true ATT remains unidentified. In this particular case, the problem reduces to the one explored in earlier misclassification papers. Section \ref{did_linear} characterizes the bias in the coefficient on the treatment using the simple DID regression without covariates and explains why a standard instrumental variables strategy will not work. Section \ref{did_x} presents our main linear conditional specification derived using the potential outcome primitives to arrive at flexible 2$\times$2 panel and repeated cross section regressions. We maintain the assumption of conditional parallel trends and derive a more general bias expression for the case with covariates. Section \ref{sol} presents the partial observability probit for one-sided misclassification and then establishes consistency and asymptotic normality of the two step FD and POLS DID estimators of the true ATT. Section \ref{app} applies the proposed method to correct for one-sided mistargeting in estimating the welfare impact of the Public Distribution System in India which suffers from targeting errors. Finally, section \ref{end} concludes. 

\section{Related Literature}\label{lit}

One of the earliest treatments of the effects of misclassified independent variables in least squares regression framework is found in \citet{aigner1973regression}. \citet{bollinger1996bounding} studies the problem of identification with a binary mismeasured regressor and provides a method to bound the model parameters. While it is commonly understood that instrumental variables can be used in dealing with errors in variables, they produce biased parameter estimates when the mismeasured variable is binary \citep{loewenstein1997delayed,black2000bounding,kane1999estimating}. However, \citet{black2003measurement} show that, in the case of non-classical measurement errors, OLS and IV estimates can bound the true parameter. \citet{kane1999estimating} and \citet{black2003measurement} show that when two mismeasured reports on the variable of interest are available, point estimates can be recovered using the GMM framework. \citet{frazis2003estimating} study the problem of misclassified and possibly endogenous regressors in linear regressions and propose a GMM based method to bound parameter estimates. Likewise, \citet{mahajan2006identification} studies identification with misclassification but relaxes the assumption of independence between measurement error and other explanatory variables. \citet{mahajan2006identification} proposes a solution akin to an IV strategy where marginal effects are identified using an additional variable correlated with the true value but uncorrelated with the measurement error. \citet{hu2008identification} generalizes Mahajan's results to multivalued discrete variables. \citet{molinari2008partial} derives partial identification results for discrete probability distributions under misclassification. \citet{kreider2010regression} studies identification of regression coefficients in a linear probability framework when a binary regressor is fully arbitrarily misclassified.

While the literature above relates to misclassification in binary explanatory variables, its real relevance is reflected in the program evaluation literature where the focus is on the parameter estimate of the treatment indicator. \citet{lewbel2007estimation} studies the identification and estimation of treatment effect with a misclassified treatment. He shows that under misclassification, the treatment effect has an attenuation bias and proposes an IV to overcome this bias. \citet{millimet2011elephant}, using Monte Carlo methods, compares the performance of different treatment effect estimators with measurement errors in treatment assignment and finds most estimators performing poorly even with minor and infrequent errors. 

In contrast to point identification approaches relying on repeated measurements or IV-type variables, \citet{kreider2012identifying} propose partial identification methods to bound the average treatment effect when the participation is misreported and possibly endogenous using semiparametric and nonparametric approaches. \citet{nguimkeu2019estimation} study point identification of treatment effects in parametric settings when participation is endogenously misreported. They propose a two-step estimation procedure that relies on \citet{poirier1980partial} partial observability model for consistent estimation of the conditional average treatment effect.

In the context of the Local Average Treatment Effect (LATE) with missing or mismeasured treatment, studies have proposed methods for both point and partial identification. When two different measurements or an instrument for the misclassified treatment is available, studies have proposed methods to point identify the LATE \citep{battistin2014misreported,ditraglia2019identifying}. \citet{yanagi2019inference} shows that LATE can be point identified under misclassified treatment if exogenous covariates are available. \citet{calvietal2021} show that point identification of LATE is possible when the treatment status of some observation is missing if two mismeasured treatment indicators are available. Moreover, under misclassification, their proposed estimator has less bias than the standard LATE estimator. \citet{ura2018heterogeneous} proposes a binary instrument for set identification in a nonparametric heterogeneous treatment effect framework with mismeasured and endogenous treatment assignment. Likewise, an IV strategy is also proposed by \citet{tommasi2020bounding} to bound the LATE with misreported survey data. \citet{acerenza2021marginal} bound the MTE under misclassification.

Although misclassified treatment in DID settings is unexplored, papers have looked at the impact of missing or misclassified treatment in longitudinal and repeated cross sectional settings. \citet{card1996effect} studies the impact of labor unions on wages in longitudinal data while explicitly accounting for misclassification errors in reported union status. \citet{botosaru2018difference} propose a method that identifies the Average Treatment Effect on the Treated (ATT) in a DID framework when individual group membership is missing in one of the time periods. Identification in their setup is achieved via the existence of proxy variables correlated with the treatment status. \citet{byker2016treatment} evaluate the impact of a female sterilization campaign implemented in Peru between 1996 and 1997 using an inverse probability weighting (IPW) estimator that accounts for contamination in the self reported sterilization status. \citet{de2018fuzzy} study identification in fuzzy DID when the share of treated units increases more in some groups than in others between the two time periods. As observed by \citet{botosaru2018difference}, \citet{de2018fuzzy} framework can also be applied to the case of missing treatment status in the baseline period.

\section{DID, Parallel Trends, and Misclassification}\label{did_simple}
To motivate the problem of misclassification, let $D^\ast$ denote the true but unobserved treatment status. However, suppose only we observe its proxy $D$ which falsely records receipt or non-receipt of the program thereby resulting in classification errors. We can write
\begin{equation}\label{mis1}
	D = D^\ast+\varepsilon
\end{equation} where by definition $\varepsilon = D-D^\ast$. As illustrated in \citet{acerenza2021marginal}, this is a general formulation and includes both; errors of inclusion and exclusion. \footnote{The triplet $(D, D^\ast, \varepsilon)$ can take any of the following values $\{(0,0,0), (1,0,1), (0,1, -1), (1,1,0)\}$. Then, $D = D^\ast \text{ if } \varepsilon \in \{0\} \text{ and }  D= 1-D^\ast \text{ if } \varepsilon \in \{1, -1\}
$} Let $S$ be a binary indicator representing the event $\varepsilon \in \{0\}$ (no misclassification). Then, we can also express (\ref{mis1}) as
\begin{equation}\label{mis2}
	D = D^\ast \cdot S+ (1-D^\ast)\cdot (1-S)
\end{equation}
Let $\{Y_t(0), Y_t(1)\}$ denote two potential outcomes (PO) where $Y_t(0)$ is the outcome in the control state and $Y_t(1)$ is the outcome in the treated state. Let $t=0$ represent the baseline period and $t=1$ represent the post-treatment period.
Since $D^\ast=0$ for everyone in the baseline, we can write the observed outcome as
\begin{equation}\label{po_mis}
	Y_0 = Y_0(0) \text{ and } Y_1 = Y_1(1)\cdot D^\ast+Y_1(0)\cdot (1-D^\ast)
\end{equation}
We assume that we either have a two-period panel or two repeated cross sections. Following \citet{wooldridgetwfe}, we can decompose each PO into 
\begin{equation}\label{po_decom}
	Y_t(d) = Y_0(d)+G_t(d)
\end{equation} which is a sum of the outcome at baseline and the gain over time, defined as $G_t(d) = Y_t(d)-Y_0(d)$ for each $d=0,1$. Therefore if $D^\ast$ were accurately observed, the DID estimand could be causally interpreted as the true $\text{ATT}$, defined as, 
\begin{equation}\label{att_true}
	\text{ATT}\equiv\mathbb{E}[Y_1(1)-Y_1(0)|D^\ast=1]
\end{equation} under the identifying assumption of unconditional parallel trends which says that \[\mathrm{DT}(D^\ast) \equiv \mathbb{E}[G_1(0)|D^\ast=1] -\mathbb{E}[G_1(0)|D^\ast=0] = 0.\]
A crucial aspect of the above identification argument is that the treatment status, $D^\ast$, is measured accurately. In other words, we know individuals' group membership into treatment and control before conducting the DID analysis. However, with a misclassified $D$, 
\begin{equation}\label{did_mis}
	\begin{split}
	\text{DID}(D) &\equiv \mathbb{E}(Y_1|D=1)-\mathbb{E}(Y_1|D=0)-\mathbb{E}(Y_0|D=1)+\mathbb{E}(Y_0|D=0) \\
	& = \underbrace{ \mathbb{E}[Y_1(1)-Y_1(0)|D=1]}_{\circled{1}} +\underbrace{\mathrm{DT}(D)}_{\circled{2}}
	\end{split}
\end{equation} where $\circled{1}$ is the ATT for those misclassified as being treated and $\circled{2}$ is the difference in counterfactual trends when we use the misclassified $D$. Therefore, even if somehow $\mathrm{DT}(D) = 0 $ which would imply parallel trends in $D$,  $\circled{1}\neq \text{ATT}$. The following proposition presents a relationship between the quantities that are estimable using a misclassified $D$ vs. $D^\ast$ under the assumption of non-differential misclassification.  
\begin{proposition}\label{prop1} In a $2\times2$ design, if we assume \textbf{non-differential misclassification} in the mean i.e. $
		\mathbb{E}[Y_1(1)-Y_1(0)|D, D^\ast] = \mathbb{E}[Y_1(1)-Y_1(0)|D^\ast]$ then
		\begin{equation*}
				\circled{1}= \mathrm{ATT}\cdot \mathbb{P}(D^\ast=1|D=1)+\mathrm{ATU}\cdot \mathbb{P}(D^\ast=0|D=1) 
		\end{equation*} where $\mathrm{ATU}$ refers to average treatment effect for those with $D^\ast=0$ (truly untreated). If in addition we assume \textbf{non-differential misclassification in counterfactual trends}, i.e.
		$\mathbb{E}[G_1(0)|D, D^\ast] = \mathbb{E}[G_1(0)|D^\ast]$ then, 
		\begin{equation*}
			\circled{2} =  \mathrm{DT}(D^\ast)\cdot\left[\mathbb{P}(D^\ast=1|D=1)+\mathbb{P}(D^\ast=0|D=0)-1\right]
		\end{equation*}
\end{proposition}
Therefore, if one is willing to assume non-differential misclassification (which may be stronger than needed), one can demonstrate that the first component of $\text{DID}(D)$ in (\ref{did_mis}) is actually a weighted average of the true ATT, true ATU, and the misclassification probabilities. 

Under non-differential misclassification in counterfactual trends, the above relationship also  implies that if parallel trends hold with the true but unobserved $D^\ast$ i.e. $\text{DT}(D^\ast)=0$, then they will also hold with the misclassified $D$ and vice versa unless $\mathbb{P}(D^\ast=1|D=1)=\mathbb{P}(D^\ast=1|D=0)$ which means that the probability of being misclassified is the same as that of being correctly classified (which is hard to justify). Note that even if $\text{DT}(D)=0$, with $D$ one can only identify the ATT for those misclassified as being treated and not the true ATT as defined in (\ref{att_true}).

\section{Bias under Misclassification with Linear CEFs}\label{did_linear}
Using (\ref{po_mis}) and (\ref{po_decom}), we can express the observed outcome in the post-treatment period as
{\small \begin{equation}\label{y1}
			Y_1 = Y_0(0)+G_1(0)+D^\ast \cdot (Y_1(1)-Y_1(0)) 
\end{equation}} The conditional mean of $Y_t$ given $D^\ast$ is 
{\small \begin{align}\label{cef}
		\mathbb{E}[Y_0|D^\ast]  =  \mathbb{E}[Y_0(0)|D^\ast] \ \text{ and } \ 
		\mathbb{E}[Y_1|D^\ast]	& = \mathbb{E}[Y_0(0)|D^\ast]+\mathbb{E}[G_1(0)|D^\ast]+ D^\ast\cdot \tau
\end{align}} where the second equality in (\ref{cef}) follows from two facts: 1) the decomposition given in (\ref{po_decom}) and 2) the simple characterization that $\mathbb{E}[Y_1(1)-Y_1(0)|D^\ast] = D^\ast \cdot \mathbb{E}[Y_1(1)-Y_1(0)|D^\ast=1]+(1-D^\ast)\cdot \mathbb{E}[Y_1(1)-Y_1(0)|D^\ast =0]$ which implies $D^\ast\cdot \mathbb{E}[Y_1(1)-Y_1(0)|D^\ast] = D^\ast \cdot \mathbb{E}[Y_1(1)-Y_1(0)|D^\ast=1] = D^\ast \tau$.   Since $D^\ast$ is binary, we can always express
\begin{equation}\label{t0_sec2}
	\mathbb{E}[Y_0(0)|D^\ast] = \eta_{0}+\eta_1\cdot D^\ast
\end{equation}
\begin{assumption}[\textbf{Unconditional parallel trends in $D^\ast$}]\label{upt} The trends in counterfactual outcomes in the absence of treatment between the truly treated and non-treated would not depend on $D^\ast$ i.e.
$\mathrm{DT}(D^\ast) = 0$.
\end{assumption} 
Parallel trends also imply that $\mathbb{E}[G_1(0)|D^\ast] = \mathbb{E}[G_1(0)] \equiv \theta$. Then combining equation (\ref{cef}) and (\ref{t0_sec2}) along with parallel trends, we obtain the simple DID equation\footnote{A similar equation can be derived for the repeated cross section setting by pooling observations for the two time periods and using the fact that the observed outcome is given by $Y = Y_0\cdot(1-T)+Y_1\cdot T$ where $T$ is a binary indicator for the post-treatment period. }
\begin{equation}\label{sdid_true}
	Y_t = \eta_0+\eta_1\cdot D^\ast+\theta\cdot t+\tau\cdot  D^\ast \cdot t+\xi_t, \ t=0,1
\end{equation} One can consistently estimate $\tau$ as the coefficient on $D^\ast$ from the first-differenced regression of $\Delta Y_i \text{ on } 1, D^\ast_i$ using a random sample of size $N$ from the population.

However, in the presence of misclassification, the coefficient on $D$ (say $\tau_{mis}$) is given by
{\small \begin{equation*}
		\begin{split}
			\tau_{mis}	= \underbrace{\tau}_{\text{truth}}\cdot \left[\underbrace{\mathbb{P}(D^\ast=1|D=1)-\mathbb{P}(D^\ast=1|D=0)}_{\text{misclassification probabilities}}\right]+\left(\underbrace{\mathbb{E}(\Delta\xi|D=1)-\mathbb{E}(\Delta\xi|D=0)}_{\text{Difference in counterfactual trends using $D$}}\right)
		\end{split}
\end{equation*}}
In the context of the simple DID regression, we see that the inconsistency arises due to two reasons. The first is an attenuation bias as seen most notably in \citet{lewbel2007estimation}, \citet{aigner1973regression}, and \citet{battistin2011misclassified} among others. The other part of the bias is due to counterfactual trends in $D$ not necessarily being parallel even though assumption (\ref{upt}) is assumed to hold. This is because misclassification may lead to counterfactual trends diverging between the $D=0$ and $D=1$ groups even when parallel trends hold with the true $D^\ast$. Our solution allows for such endogenous or differential misclassification.\footnote{If, however, we assume  non-differential misclassification between $S$ and counterfactual trends then a simple relationship between $\text{DT}(D)$ and $\text{DT}(D^\ast)$ emerges, as seen in section \ref{did_simple}.}

\subsection{Would Instrumental Variables Work?}
Can the instrumental variables approach help us in obtaining a consistent estimator of $\tau$? To see this, let's suppose we have a scalar instrument, $Z$, which is relevant for $D$. In addition to that, the IV also needs to be exogeneous to the regression error. Consider the first-dfferenced equation 
\begin{equation*}
 	\Delta Y_i = \theta+\tau D_i +\Delta \epsilon_i, 
\end{equation*}
where the IV must satisfy $	\mathrm{Cov}(Z, \Delta \epsilon) = 0$. However, 
\begin{equation*}
	\mathrm{Cov}(Z, \Delta \epsilon)= \mathrm{Cov}[Z, \tau(D^\ast-D)+\Delta \xi] = \tau\cdot \mathrm{Cov}[Z, (D^\ast-D)]+\mathrm{Cov}(Z, \Delta \xi) 
\end{equation*} 
Being truly relevant means that $Z$ is also correlated with the misclassification error. Therefore the first covariance will be non-zero.  This is because measurement error in a binary variable necessarily implies that the truth is negatively correlated with the error. Second, since misclassification may be correlated with $\Delta \xi$, $\mathrm{Cov}(Z, \Delta\xi)\neq 0$. Hence, a simple IV strategy will produce a biased estimator for $\tau$.

\section{DID with Covariates}\label{did_x}
Often parallel trends are only plausible once we allow selection into treatment to be based on observables. To allow for this, we let $X = (X_1, X_2, \ldots, X_k)$ denote a vector of pre-treatment covariates.\footnote{For ease of derivations later on, we will use $R = (1, X)$ to be the set which includes an intercept.} Suppose that  conditional parallel trends hold. Formally, 
\begin{assumption}[\textbf{Conditional parallel trends}]\label{as:cpt} $\mathbb{E}[G_1(0)| X, D^\ast] = \mathbb{E}[G_1(0))|X]$	
\end{assumption} 
Assumption (\ref{as:cpt}) allows for covariate specific time trends in the potential outcome mean for the treated and control groups. DID studies that work with this assumption include \citet{abadie2005semiparametric}, \citet{callaway2020difference}, \citet{sant2020doubly}, \citet{wooldridgetwfe} etc. 

\begin{assumption}[\textbf{Overlap}]\label{as:overlap} $\mathbb{P}(D^\ast=1) >0$ and  $\mathbb{P}(D^\ast=1|X)<1$.	
\end{assumption} Assumption \ref{as:overlap} is the overlap condition required for the identification of the ATT.  

We also assume that the researcher has access to either a two-period panel or two repeated cross sections. This is formalized in the assumption below.
\begin{assumption}[\textbf{Random sampling scheme}]\label{as:rs} Assume that the data are either independent and identically distributed from 
	i) Panel:  $\{(Y_{it},X_i,D^\ast_i); \ i=1,2,\ldots,N\}$, for $t=0,1$ constitues an \textit{i.i.d} draw from the population; ii) Repeated cross section: Conditional on $T=0$, the data are \textit{i.i.d} from the distribution $\{(Y_{0}, D^\ast, X);\}$; conditional on $T = 1$, the data are \textit{i.i.d}. from the distribution $\{(Y_{1}, D^\ast, X)\}$. 
	\begin{equation*}
		\begin{split}
			\mathbb{P}(Y\leq y,D^\ast=d,X\leq x,T=t) &= t\cdot p\cdot P(Y_1\leq y,D^\ast=d,X\leq x|T=1) \\ &+(1-t)\cdot (1-p)\cdot P(Y_0 \leq y,D^\ast = d,X\leq x|T = 0),
		\end{split}
	\end{equation*}
	where $\lambda \equiv \mathbb{P}(T=1)\in (0,1)$ and $(y,d,x,t) \in \mathbb{R}\times\{0,1\}\times \mathbb{R}^k\times \{0,1\}$.
\end{assumption}
Assumption \ref{as:rs} i) discusses the random sampling scheme in the case of a two-period panel where we have repeated observations on individuals for both the pre and post treatment periods. Part ii) discusses the sampling scheme in the case of two repeated cross sections where we observe each individual only once, either in the pre-treatment or the post-treatment period. This assumption rules out compositional changes in the sample (see \citet{hong2013measuring}). 

In the presence of covariates the conditional expectations, $\mathbb{E}[Y_t|X, D^\ast]$, are given as 
 \begin{equation}\label{cefx}
 	\mathbb{E}[Y_0|X, D^\ast] = \mathbb{E}[Y_0(0)|X, D^\ast] \text{ and } 	\mathbb{E}[Y_1|X, D^\ast]= \mathbb{E}[Y_1(0)|X, D^\ast] + D^\ast\cdot \tau(X)
 	\end{equation} where $\tau(X) = \mathbb{E}[Y_1(1)-Y_1(0)|X, D^\ast=1]$. 
Assume that
\begin{align}
	\mathbb{E}[Y_0(0)|X, D^\ast] & = \eta_{00}+\eta_{01}D^\ast+\mathring{X}\eta_{02}+D^\ast\mathring{X}\eta_{03} \label{lip1} \\
	\mathbb{E}[G_1(0)|X] &= \delta_{01}+\mathring{X}\delta_{11} \label{lip2} \\
	\tau(X) &= \tau+\mathring{X}\kappa \label{lip3}
\end{align}
where $\mathring{X} = X-\mathbb{E}(X|D^\ast=1)$. Then substituting (\ref{lip1}), (\ref{lip2}), (\ref{lip3}) in (\ref{cefx}), we can generally express
\begin{equation}\label{ey}
	\begin{split}
		Y_t &= \mathring{R}\eta_1+W^\ast\eta_2+t\cdot \mathring{R}\delta+t \cdot W^\ast\theta+\xi_t \ \text{ such that } \\
		&\mathbb{E}(\xi_t) = 0, \ \  \mathbb{E}[\xi_t|X, D^\ast] = 0 
	\end{split}
\end{equation} where $\mathring{R} \equiv (1, \mathring{X})$, $W^\ast \equiv  D^\ast \mathring{R}$, $\delta = (\delta_{01}, \delta_{11}^\prime)^\prime$, $\eta_1 = (\eta_{00}, \eta_{02}^\prime)^\prime$, $\eta_2 = (\eta_{01}, \eta_{03}^\prime)^\prime$, and $\theta = (\tau, \kappa)^\prime$.
Note that our framework allows the covariates to affect the dynamics of the potential outcomes and also considers the treatment effect to be heterogeneous based on covariates.
With a two-period panel, one can estimate the first-differenced equation 
\begin{equation}\label{fd_ifreg}
	\begin{split}
		&\Delta Y_i = \mathring{R}_i\delta+W_i^\ast\theta+\Delta\xi_i, \ i=1,\ldots, N \\
		&\text{ where } \ \ \Delta\xi_i = \xi_{i1}-\xi_{i0}
	\end{split}
\end{equation}

\paragraph{Repeated cross sections} Often DID analysis is also carried out using two repeated cross sections since obtaining data on same observations before and after the treatment is not feasible. In that case, one can estimate the pooled regression 
\begin{equation}\label{rc_ifreg}
	\begin{split}
		Y_i &=  \mathring{R}_i\eta_1+W_i^\ast\eta_2+T_i\cdot \mathring{R}_i\delta+T_i\cdot W_i^\ast\theta+\xi_i, \ i=1,\ldots, N \\
		&\text{ where } \ \   \xi_i= \xi_{i1}\cdot T_i+\xi_{0i}\cdot (1-T_i)
	\end{split}
\end{equation} Note that the true treatment indicator, $D^\ast$, and covariates, $X$, are exogenous in equations (\ref{fd_ifreg}) and (\ref{rc_ifreg}) above. Therefore, if we perfectly observe $D^\ast$, we can consistently estimate $\theta$, and consequently $\tau$, as the coefficient on $D^\ast$ from the equations above depending on the kind of sample we have.

\subsection{Partial Observability Probit}\label{onesided} 
In the current paper, we propose a solution for only one-sided misclassification by assuming that 
\begin{equation}
	D = D^\ast \cdot S
\end{equation} which means that $D=1$ only if $D^\ast=1$ and $S=1$. Apart from this, the one-sided formulation is only able to capture errors of exclusion where $D^\ast=1$ but $D=0$  which corresponds to $S=0$. Errors of inclusion (i.e. $D=1$ when $D^\ast=0$) are not explicitly accounted for in equation (\ref{onesided}).   \label{panel_miss}	
Suppose, 
\begin{equation*}
	D^\ast= \mathbbm{1}\{ R\gamma+U\geq 0\}; \text{ } S = \mathbbm{1}\{Z\alpha +V\geq 0\}
\end{equation*}
Then, 
\begin{equation}\label{pop}
	D = \mathbbm{1}\{R\gamma+U\geq 0, Z\alpha +V\geq 0\}
\end{equation} Let the conditional distribution of $(-U, -V)$ be bivariate normal with the CDF denoted by $F_{U, V}(\cdot, \cdot; \rho)$ where $\rho$ is the correlation coefficient (see assumption \ref{as:errors}). Then, 
\begin{equation*}
\mathbb{P}(D=1|R, Z) = \mathbb{P}(-U\leq R\gamma, -V\leq Z\alpha) = F_{U,V}(R\gamma, Z\alpha; \rho)
\end{equation*}
Identification of parameters $\alpha$ and $\gamma$ in the partial observability probit model requires one exogenous variable in $Z$ that is excluded from $R$ in order to satisfy local identification assumptions in \citet{poirier1980partial}. 

The DID design allows for the treatment, $D^\ast$, to be endogenous only due to time invariant unobservables. Therefore, comparisons between treated and non-treated groups in the baseline and post-treatment periods provide a valid estimate of the ATT.  The design, therefore, dictates a clear relationship between the outcome error and $U$. With respect to misclassification, since $S$ may be correlated with time varying unobservables affecting the outcome dynamics, we allow $V$ and the outcome error to be correlated. We already mention that this will lead to parallel trends not holding with $D$ even if we assume them to hold with $D^\ast$.\footnote{except when misclassification depends only on observables in which case we will obtain a similar result as section \ref{did_simple} conditional on $X$.} We modify the assumptions in \citet{nguimkeu2019estimation} to reflect these subtleties. 

\begin{assumption}\label{as:errors} Assume that 
	\begin{itemize}
		\item[a.] The error term $\Delta\xi$ is independent of $R$, $Z$, with variance $\sigma^2$; and the error terms $(U,V)$ are independent of all covariates $R$, $Z$ and have unit variances. The correlations for the pairs $(\Delta \xi,V)$ and $(U, V)$ are denoted $\psi_v$ and $\rho$, respectively.
		\item[b.] The error terms $(\Delta\xi, U, V)$, follow a trivariate normal distribution, conditional on all covariates $(R,Z)$.
		\begin{equation*}
			(\Delta\xi, U, V)^\prime|R, Z \sim N(0,\Sigma) \text{ where } \Sigma = \begin{pmatrix}
				\sigma^2 & 0 & \psi_v\sigma \\
				0 & 1 & \rho \\
				\psi_v\sigma & \rho & 1
			\end{pmatrix} 
		\end{equation*}
	\end{itemize}
In the case of repeated cross sections, simply replace $\Delta \xi$ with the pooled error $\xi$. 
\end{assumption}
Given that we observe $D$, the first differenced equation becomes,  
\begin{equation}\label{fd_freg}
	\begin{split}
		&\Delta Y_i = \ddot{R}_i\delta+\ddot{W}_i\theta+\Delta \epsilon_i, \ \ i=1,\ldots, N \\
		\Delta \epsilon_i &=[(\mathring{R}_i-\ddot{R}_i)\delta+(W_i^\ast-\ddot{W}_i)\theta+\Delta \xi_i]
	\end{split}
\end{equation} 
Note that $\ddot{R}_i \equiv (1, \ddot{X}_i)$ where $\ddot{X}_i = X_i-\bar{X}_1$ with $\bar{X}_1$ being the sample mean of covariates for those misclassified as being treated $(D_i=1)$. Similarly, $\ddot{W}_i \equiv D_i\ddot{R}_i$. Define $\hat{\theta}_{FD}$ to be the estimator of the coefficient on $\ddot{W}_i$ from estimating the FD equation given in (\ref{fd_freg}).

\paragraph{Repeated cross-sections}
In the case of repeated cross sections, we obtain the following pooled regression equation
\begin{equation}\label{pols_freg}
			Y_i =  \ddot{R}_i\eta_1+\ddot{W}_i\eta_2+T_i\cdot \ddot{R}_i \delta + T_i\cdot \ddot{W}_i\theta+\epsilon_i, \ i=1,\ldots, N
\end{equation} where $\epsilon_i = \left[ (\mathring{R}_i-\ddot{R}_i)\eta_1+(W_i^\ast-\ddot{W}_i)\eta_2+T_i\cdot (\mathring{R}_i-\ddot{R}_i)\delta+T_i\cdot (W_i^\ast-\ddot{W}_i)\theta+\xi_i\right]$. Define $\hat{\theta}_{POLS}$ to be the estimator for $\theta$ from estimating equation (\ref{pols_freg}) using a pooled sample. 
	
Since $D$ is endogenous to both equations, $\hat{\theta}_{FD}$ and $\hat{\theta}_{POLS}$ will be inconsistent for $\theta$. We characterize this asymptotic bias from using a misclassified $D$ in place of $D^\ast$. 
\begin{theorem}[Bias under misclassification with covariates]\label{thm:bias_panel} Under assumptions \ref{as:cpt}, \ref{as:overlap}, \ref{as:rs}, and \ref{as:errors} 
	\begin{equation*}
			\mathrm{plim}(\hat{\tau}_{FD})-\tau = \mathbb{E}[\dot{R}_i  Q^{-1}(A\delta+B\theta+C)|D_i=1] 
	\end{equation*}
	where {\small $Q =  \mathbb{E}(\dot{W}_i^\prime 		\dot{W}_i)-\mathbb{E}(\dot{W}_i^\prime\dot{R}_i)\mathbb{E}(\dot{R}_i^\prime\dot{R}_i)^{-1}\mathbb{E}(\dot{R}_i^\prime\dot{W}_i)$, $A =  \mathbb{E}(\dot{W}_i^\prime\mathring{R}_i)- 		\mathbb{E}(\dot{W}_i^\prime\dot{R}_i)[\mathbb{E}(\dot{R}_i^\prime\dot{R}_i)]^{-1}\mathbb{E}(\dot{R}_i^\prime\mathring{R}_i) $, $	B = \mathbb{E}[\dot{W}_i^\prime(W_i^\ast-\dot{W}_i)] - 		\mathbb{E}(\dot{W}_i^\prime\dot{R}_i)[\mathbb{E}(\dot{R}_i^\prime \dot{R}_i)]^{-1}\mathbb{E}[\dot{R}_i^\prime(W_i^\ast-\dot{W}_i)$ and $C = \sigma\psi_v\mathbb{E}\left[\dot{R}_i^\prime \phi(-Z_i\alpha)\Phi\left(\frac{R_i\gamma-\rho Z_i\alpha}{\sqrt{1-\rho^2}}\right)\right] $} and 
	\begin{equation*}
		\mathrm{plim}(\hat{\tau}_{POLS})-\tau = \mathbb{E}[\dot{R}_i  \left\{Q_1^{-1}(A_1\pi_1+B\pi_2+C_1)-Q_0^{-1}(A_1\eta_1+B\eta_2+C_0)\right\}|D_i=1]
	\end{equation*}
where $Q, A, B$ and $C$ are defined for the $T=1$ and $T=0$ populations. 
\end{theorem} 
As we can see from the above theorem, the asymptotic bias is not  a simple attenuation bias and allows for endogeneous misclassification in $D$.

\section{Two-Step Solution}\label{sol}
The first step of the two-step solution we propose estimates the partial observability probit parameters $(\gamma, \alpha, \rho)$ by maximizing the log-likelihood function given by
\begin{equation*}
	\underset{(\gamma, \alpha, \rho)}{\mathrm{max}}\sum_{i=1}^{N}D_i\mathrm{log}[F_{U,V}(R_i\gamma, Z_i\alpha; \rho)]+(1-D_i)\mathrm{log}[1-F_{U,V}(R_i\gamma, Z_i\alpha; \rho)]
\end{equation*}
We then use the first step maximum likelihood estimate of $\alpha$ to obtain a predicted $D_i^\ast$ as $\hat{D}_i^\ast = \Phi(R_i\hat{\gamma})$  and plug it in true regression equations. The first-differenced equation becomes
\begin{equation}\label{freg2}
	\begin{split}
		\Delta Y_i &= \hat{R}_i^\ast\delta+\hat{W}_i^\ast\theta+\Delta\varepsilon_i, \ i=1, \ldots, N \\
		\Delta\varepsilon_i&= (\mathring{R}_i-\hat{R}_i^\ast)\delta+(W_i^\ast-\hat{W}_i^\ast)\theta+\Delta\xi_i 
	\end{split}
\end{equation}	
and $\hat{R}^\ast \equiv (1, \hat{X}^\ast)$ where $\hat{X}^\ast = X-\hat{\bar{X}}_1^\ast$ and $	\hat{\bar{X}}_1^\ast = \frac{1}{\hat{N}^\ast}\sum_{i=1}^{N}\hat{D}^\ast_i\cdot X_i$ with $\hat{N}^\ast = \sum_{i=1}^{N}\hat{D}^\ast_i$ being the sum of predicted probabilities. Also, $\hat{W}^\ast \equiv \hat{D}^\ast \hat{R}^\ast$. Then, using Frisch-Waugh again, the two-step estimator will be obtained from the following regression
\begin{equation}\label{freg2m}
	\hat{M}^\ast \Delta Y = \hat{M}^\ast\hat{W}^\ast\theta+\hat{M}^\ast\Delta\varepsilon 
\end{equation} where $\hat{M}^\ast = I-\hat{R}^\ast(\hat{R}^{\ast\prime}\hat{R}^\ast)^{-1}\hat{R}^{\ast\prime}$ and all variables involved are expressed in vector and matrix notations. Then, 
\begin{equation}\label{estimator_2S}
	\hat{\theta}_{FD}^{2S} = \left[\hat{W}^{\ast\prime} \hat{M}^\ast\hat{W}^\ast\right]^{-1}\hat{W}^{\ast\prime}\hat{M}^\ast\Delta Y 
\end{equation}
where $\hat{\theta}_{FD}^{2S}$ is the proposed two-step FD estimator.

For the case of a repeated cross section, the DID estimator is obtained from equation (\ref{rc_ifreg}) by plugging in $\hat{D}^\ast$ in place of $D$ to obtain the following equation
\begin{equation}\label{freg_rc}
	\begin{split}
		Y_i =  \hat{R}_i^\ast\eta_1+\hat{W}_i^\ast\eta_2+T_i\cdot \hat{R}_i^\ast \delta + T\cdot \hat{W}_i^\ast\theta+\varepsilon_i
	\end{split}	
\end{equation} where $\varepsilon_i = \left[ (\mathring{R}_i-\hat{R}^\ast_i)\eta_1+(W_i^\ast-\hat{W}^\ast_i)\eta_2+T\cdot (\mathring{R}_i-\hat{R}^\ast_i)\delta+T\cdot (W_i^\ast-\hat{W}^\ast_i)\theta+\xi_i\right]$. 

Next, we establish consistency and asymptotic normality (along with the variance-covariance expressions) of the two-step FD and POLS DID estimators. 

\begin{theorem}[Asymptotic distribution of the two-step estimator]\label{thm:asn_panel} Under assumptions \ref{as:cpt}, \ref{as:overlap}, \ref{as:rs}, and \ref{as:errors} given above, 
	\begin{equation*}
		\begin{split}
		\sqrt{N}(\hat{\tau}_{FD}^{2S}-\tau) &\overset{d}{\rightarrow} N(0, \Omega_{FD}) \\
		\Omega_{FD} = \mathbb{E}(\mathring{R}_i|D_i^\ast=1)&\cdot \text{Avar}[\sqrt{N}(\hat{\theta}_{FD}^{2S}-\theta)]\cdot \mathbb{E}(\mathring{R}_i|D_i^\ast=1)^\prime
		\end{split}
	\end{equation*} where
\begin{equation*}
	 \text{Avar}[\sqrt{N}(\hat{\theta}_{FD}^{2S}-\theta)]=\Omega_{\Gamma}^{-1}\left(\Omega_{1\theta}+\Omega_{2\theta}+\Omega_{3\theta}\right)\Omega_{\Gamma}^{-1}
\end{equation*}
and 
\begin{equation*}
	\begin{split}
		\sqrt{N}(\hat{\tau}_{POLS}^{2S}-\tau) &\overset{d}{\rightarrow} N(0, \Omega_{POLS}) \\
		\Omega_{POLS}& = \mathbb{E}(\mathring{R}_i|D_i^\ast=1)\cdot \text{Avar}[\sqrt{N}(\hat{\theta}_{POLS}^{2S}-\theta)]\cdot \mathbb{E}(\mathring{R}_i|D_i^\ast=1)^\prime \\
		\text{Avar}[\sqrt{N}(\hat{\theta}_{POLS}^{2S}-\theta)] &=\Omega_{\Gamma}^{-1}\bigg\{\left(\frac{\Omega_{1\pi}}{\lambda}+\frac{\Omega_{1\eta}}{(1-\lambda)}\right)+\left(\frac{\Omega_{2\pi}}{\lambda}+\frac{\Omega_{2\eta}}{(1-\lambda)}\right)\\
		&+\left(\frac{\Omega_{3\pi}}{\lambda}+\frac{\Omega_{3\eta}}{(1-\lambda)}\right)\bigg\}\Omega_{\Gamma}^{-1}  
	\end{split}
	\end{equation*} 
\end{theorem}
where $\pi$ are the regression parameters for the $T=1$ population and $\eta$ index the $T=0$ population regression.

\section{Empirical Application}\label{app}

We apply the proposed method to study the insurance impact of a large-scale in-kind transfer program in India. The program in question is the Public Distribution System (PDS) of India. The PDS is the world's largest targeted food safety net program that distributes highly subsidized food grains, mainly rice and wheat, to close to a billion people in 180 million poor eligible households \citep{balani2013functioning,gadenne2021kind}. Initially, the program had universal coverage, but in 1997 it was transformed into the Targeted Public Distribution System (TPDS), which emphasized targeted food subsidies for only the poor eligible households.

The TPDS differentiates between households below (BPL households) and above (APL households) the official poverty line. The state governments have the responsibility of identifying the BPL households. The state governments rely on the elected village governments who in turn identify the poor using multiple criteria, including qualitative parameters such as possession of land operated/owned; ownership of TVs, motorcycles, and other durables; and ownership of agricultural machinery and implements \citep{kochar2005can,ram2009understanding,kaushal2015consumer}. The BPL and APL households are issued different ration cards that identify them as entitled to either the BPL or the APL subsidy. The BPL households are the priority households and are allocated food grains at much lower prices than the APL households. Although any household above the poverty line is entitled to the APL ration card, the main beneficiaries of the TPDS are the households with the BPL ration card.

We are interested in estimating the insurance impact of access to cheap grains through PDS on eligible or treated households in India. \citet{negi2022global} shows that during the global food price surge in 2007-2008, Indian households with access to PDS subsidy were able to maintain their staple food consumption and total calorie intakes by substituting expensive market purchased food grains with subsidized PDS grains. \citet{gadenne2021kind} in the context of India, show that in-kind transfers provide implicit insurance to eligible households from commodity price risk. Their key argument is that in-kind transfers have insurance benefits as the value of the transfer is naturally indexed to the market value of the commodity and hence rises as the market price of the commodity rises. This implies that if food prices go up, access to in-kind transfers will allow households to maintain their real value of income. This is the key argument we empirically test in this paper. The challenge is that the PDS is known to have large targeting errors \citep{dutta2001targeting, swaminathan2001errors, hirway2003identification, khera2008access}. Moreover, there is some evidence that the errors of wrong exclusion are much larger than errors of wrong inclusion \citep{jha2013safety, pingali2019reimagining}. Estimates suggest that targeting errors continue to be significant as only 28\% in the bottom 40\% of the households access PDS \citep{pingali2019reimagining}. \citet{jha2013safety} report that the proportion of poor who used the PDS in 2004-2005 was only 30\% and the exclusion error was as high as 70\%. They go on to report that a part of this exclusion error was due to targeting errors where some of the poor were not identified as poor and hence were ineligible for PDS subsidies.

In principle, targeting should increase the efficiency of the program by transferring scarce government resources directly to groups that need them the most. However, in reality, the performance of such programs varies based on the targeting methods and the quality of information on the population \citep{coady2004targeting}. A detrimental factor influencing the efficacy of such programs is the precision with which the targeted group is identified in the population \citep{coady2004targeting}. Errors in targeting are common and arise due to imperfect information about eligibility, corruption, political connections, and elite capture \citep{cornia1993two, coady2004targeting,pande2007understanding, besley2012just, alatas2012targeting, panda2015political}. For these reasons, these programs generally suffer from \textit{errors of inclusion} or identifying non-eligible as eligible and \textit{errors of exclusion} or identifying eligible as non-eligible.
 
In general, a mistargeted program poses a challenge in terms of its impact evaluation. A routinely used method to assess the performance of targeted programs is difference-in-differences  (DID), but in the presence of targeting errors, it's not clear what the estimated treatment effect would capture. In such a scenario, the treatment effect from a DID regression will not reflect the program's true impact as the observed treatment group would be different from the intended treatment group. Although mistargeting is a reality in targeted social welfare programs, few studies have taken account of it while estimating the welfare impacts of such programs \citep[see,][for examples]{emran2014assessing,tohari2019targeting}. \citet{cameron2014can} report an extreme case from Indonesia where mistargeting of a cash transfer program actually led to destruction of trust and social capital and increase in criminal behavior.

\subsection{Data and Setting}

We use data from the Indian Human Development Surveys (IHDS) conducted in 2004-05 and 2011-12 to assess the insurance impact of PDS subsidies on Indian households \citep{desai2010national,desai2018national}. The IHDS are large scale, nationally represented panel household surveys conducted by the National Council of Applied
Economic Research (NCAER) India, the University of Maryland, Indiana University, and the University of Michigan. The IHDS collect household and individual level data on a wide variety of indicators including household income, expenditure,
assets, education, caste, gender relations, local infrastructure, and availability of facilities. More importantly, the IHDS include self reported ownership of ration cards and the type (BPL or APL) of ration cards owned. 

Table \ref{tab:sum} presents the summary statistics on key variables of interest from the baseline and endline IHDS. We categorize households with no ration card or APL card as the control group as they are not the main beneficiaries of the PDS subsidies. The treatment group is defined as households owning a BPL or an AAY (Antyodaya Ann Yojna) ration card, as they are the intended beneficiaries of the program.\footnote{AAY ration card is issued to the poorest households within the BPL category with more generous subsidies.} Since the treatment group has to be the same in both the survey rounds, we retain only those households whose ration card status remained constant in both the rounds. We also remove households in the state of Tamil Nadu as it didn't go for TPDS and followed universal PDS during the period of analysis. Overall, we are left with 3035 households in the treated group and 6345 in the control group.

In terms of income and assets, the treated households are poorer than households in the control group. This is expected as BPL ration cards are primarily given to poor households. One important control variable in Table \ref{tab:sum} is the household's participation in wage work under Mahatma Gandhi National Rural Employment Guarantee Act (MGNREGA). MGNREGA is India's large-scale anti-poverty rural workfare program which was introduced in 2005 to provide around 100 days per year of minimum wage employment to working age individuals. The MGNREGA is primarily operational in rural areas and offers voluntary unskilled employment on local public work projects. Another critical control is monetary benefits received from other government programs, calculated as the sum of transfers received from scholarships, old age pension, maternity schemes, disability schemes, income generation programs other than MGNREGA, drought/flood compensation assistance, and insurance payouts. This variable is coded as zero if no transfer was recieved; otherwise, the monetary amount of transfer. Note that participation and benefits from both of these programs show an increasing trend during this period and hence are important controls in our empirical specification. 

Interestingly, treated households report having higher membership in the local caste associations and links to local politicians than the control group. This is consistent with the observations made in the literature that elite capture and connections with local politicians is helpful in selecting beneficiaries for transfer programs in developing countries \citep{pande2007understanding,besley2012just,panda2015political}. In the context of the PDS program in India, \cite{panda2015political} shows that local political connections are conducive to being selected into getting a BPL ration card. These observations are important for us as we will use these variables to correct for the influence of mistargeting in our estimates.

\begin{figure}[H]\caption{\textbf{Rice and Wheat Prices in Global and Domestic Markets}}\label{fig:prices}
		\noindent
		\makebox[\textwidth]{\includegraphics[scale=0.12]{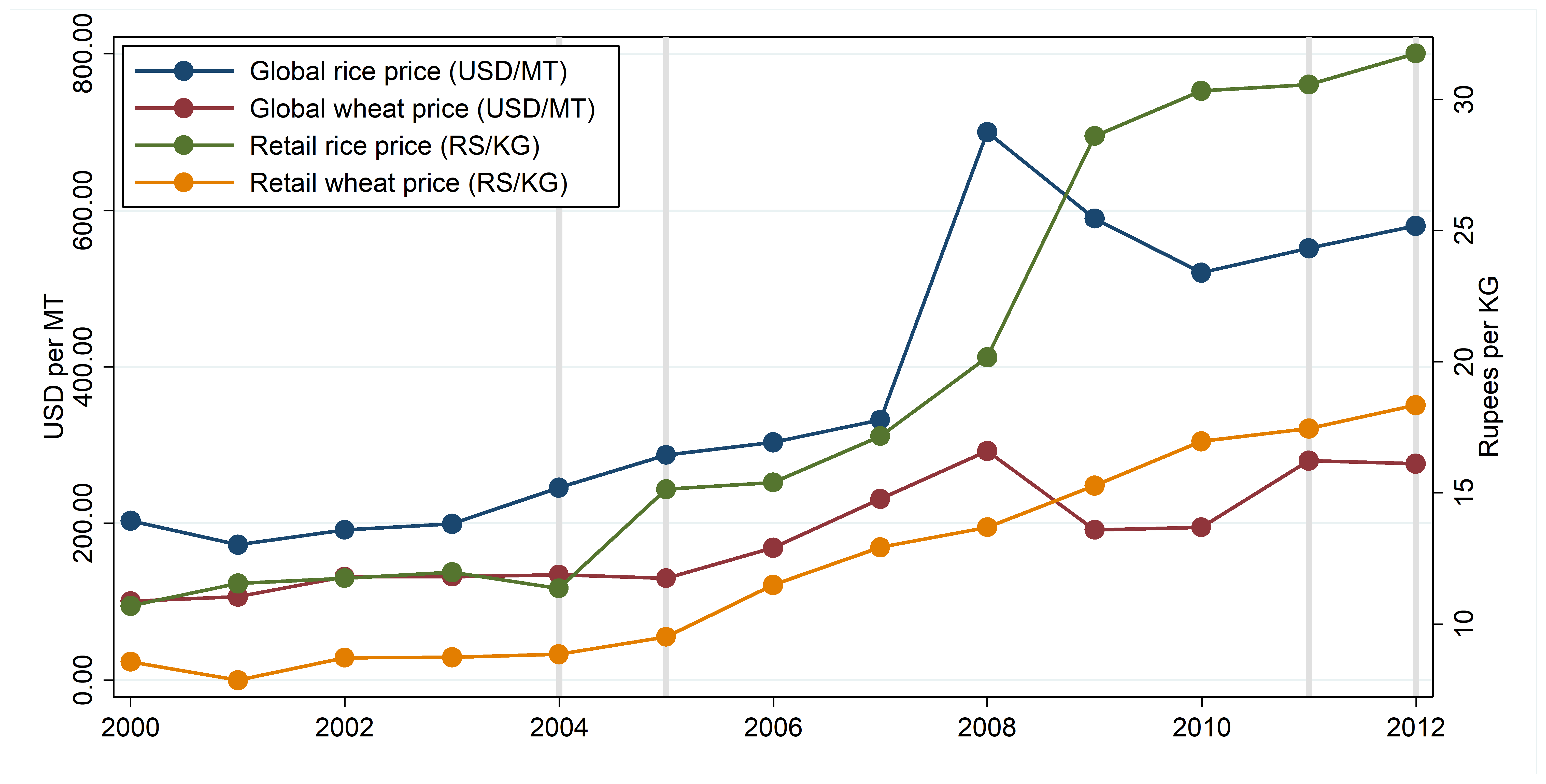}}
		\centering	\begin{minipage}{13cm}~\\
			\footnotesize  Notes: Price of Thai rice and US wheat in USD per metric ton as the global price of the two commodities. Retail prices are prices of rice and wheat prevailing in the retail shops. The retail price data comes from the price monitoring bureau of the Department of Consumer Affairs, Ministry of Agriculture, Government of India. These are nominal prices.
		\end{minipage}
	\end{figure}

\begin{figure}[H]\caption{\textbf{Market Purchased and PDS Price of Rice and Wheat}}\label{fig:ihdsprice}
	\noindent
	\makebox[\textwidth]{\includegraphics[scale=0.12]{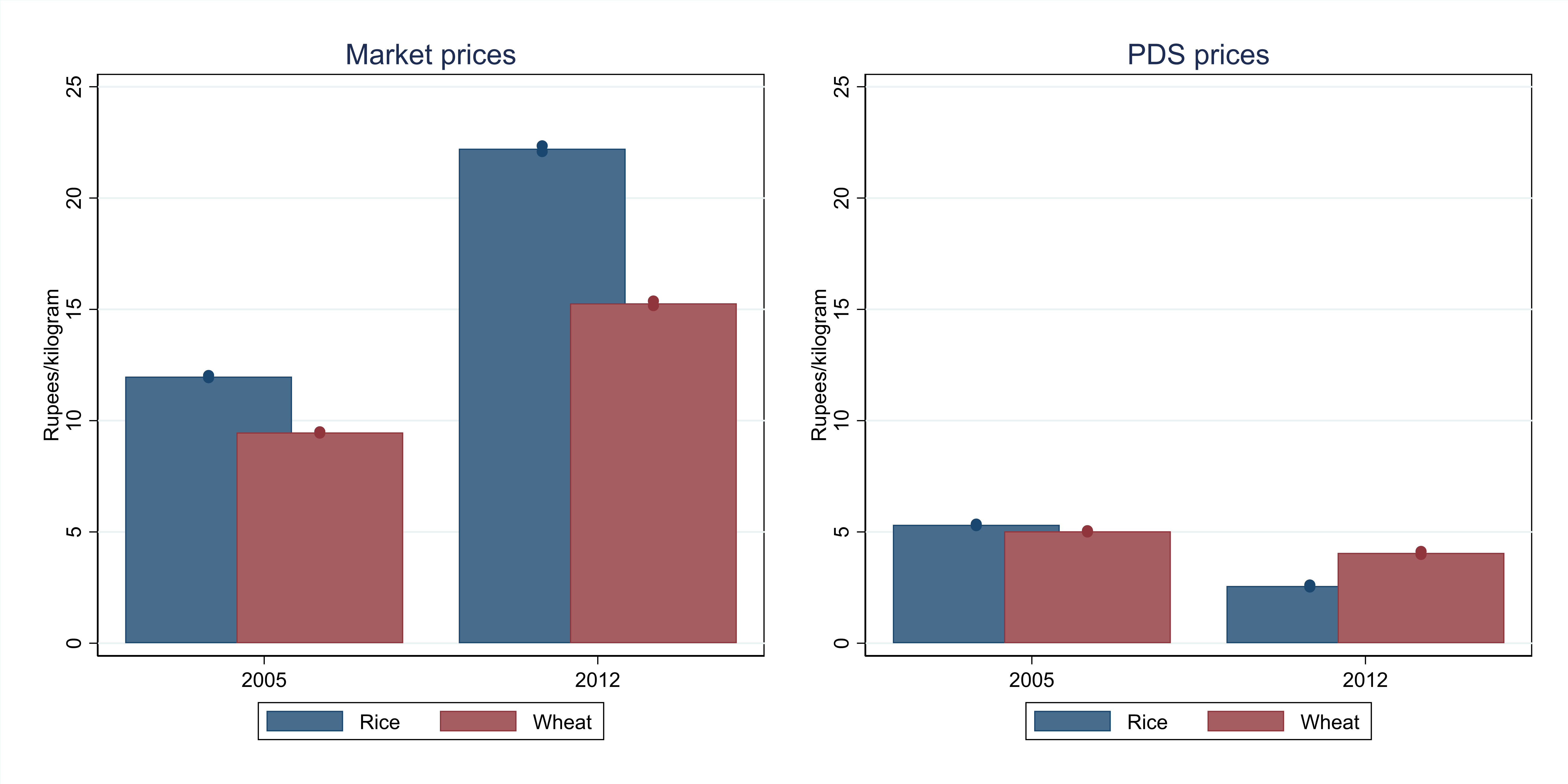}}
	\centering	\begin{minipage}{13cm}~\\
		\footnotesize  Note: Market and PDS price of rice and wheat reported in the IHDS with 95\% confidence intervals. These are nominal prices.
	\end{minipage}
\end{figure}

To set the context, we present global and domestic price trends in Figure \ref{fig:prices}. As can be seen from the figure, global and domestic rice and wheat prices were on the rise around 2007-08. Evidence from the literature suggests that global food price increase around this period was triggered by productions shocks in major producing countries and ensuing countercyclical trade policies \citep{negi2022global}. Given pressure form rising global food prices, domestic food prices in India also started trending upwards so much so that the price of rice and wheat almost doubled from their levels in 2004-05. Figure \ref{fig:ihdsprice} shows the average market and PDS price of rice and wheat estimated from the IHDS for the baseline and the endline period. The market price of rice and wheat show an increase in the IHDS as well but the PDS price registers a marginal decline. The decline is probably on account of increased subsidies on PDS rice and wheat during this period \citep{gadenne2021kind}.

Table \ref{tab:cons} shows suggestive evidence on the role of PDS in insulating eligible households' food consumption from high food prices. The total consumption of rice and wheat remains comparable for both the treated and control groups across the two rounds, but the shift from market purchased to subsidized rice and wheat is clearly visible for the treated group. In comparison, for the control group, the decline in market purchased rice and wheat is minor.   

\begin{figure}[H]\caption{\textbf{Calorie Intakes by Observed BPL Ration Card Status for Baseline and Endline}}\label{fig:cal}
	\noindent
	\makebox[\textwidth]{\includegraphics[scale=0.6]{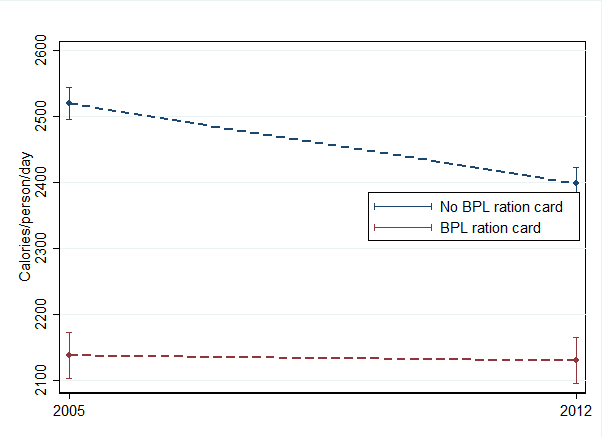}}
	\centering	\begin{minipage}{11cm}~\\
		\footnotesize  Note: Average per person per day calorie intakes for BPL ration card onwing and non owning households for the baseline and the endline survey with 95\% confidence intervals.
	\end{minipage}
\end{figure}

Our outcome of interest is the daily per capita calorie intakes which we calculate from item-wise food consumption data reported in the IHDS. We select total calories intakes as the outcome as it reflects food security and is a better measure in this context than other monetary measures of welfare. Figure \ref{fig:cal} shows the trends in the average per capita calorie intakes. Two observations are worth noting from Figure \ref{fig:cal}. First, average calorie intakes of BPL ration card owning households are lower than the that of the APL ration card owning households. This basically reflects the fact that BPL ration card households are generally poorer households. The second and more interesting observation is that the calorie intakes of the BPL ration card owning households is stable across the baseline and the endline whereas for APL ration card owning households, it shows a marked decline. Our hypothesis is that, during the period of high food prices, BPL households could maintain their calorie intakes essentially because they could access cheap grains from the PDS. APL households had either limited or no access to government subsidized grains hence could not maintain their calorie intakes. In the next section, we illustrate how we use observations from Figures \ref{fig:ihdsprice} and \ref{fig:cal} to propose a DID strategy to estimate the insurance benefits of the PDS subsidy which is mainly targeted to the BPL ration card owning households.  

\subsection{Differences-in-Differences with Targeting Errors}

\begin{figure}[H]\caption{\textbf{Visualizing the Empirical Strategy}}
	\noindent\label{fig:did}
	\makebox[\textwidth]{\includegraphics[scale=0.34]{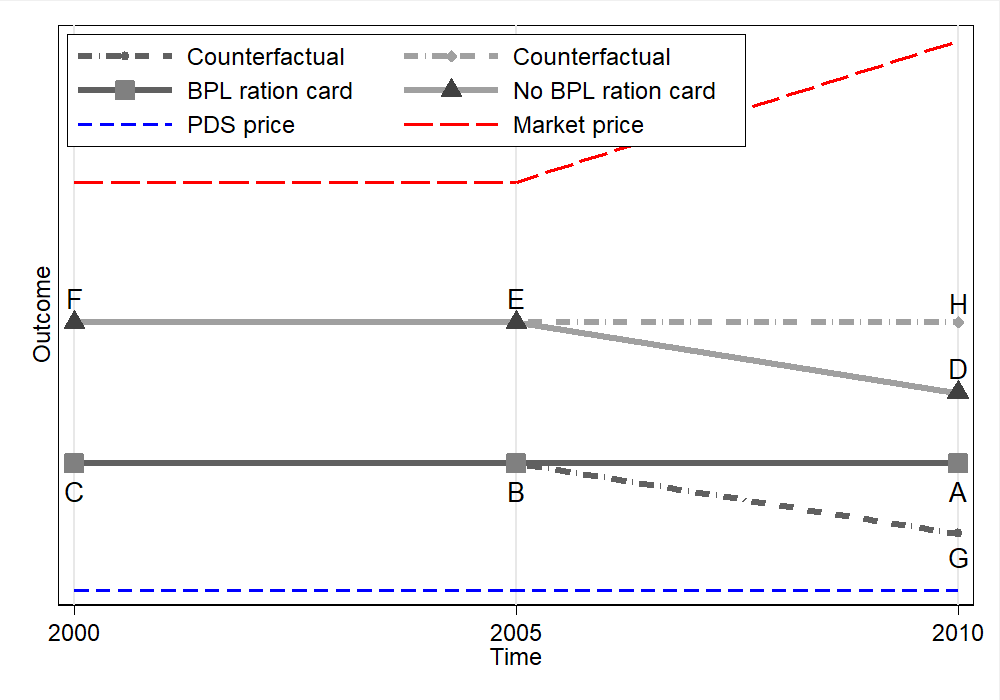}}
	\centering	\begin{minipage}{11cm}~\\
		\footnotesize  Note: Figure plots hypothetical values for outcome and prices to illustrate the empirical strategy.
	\end{minipage}
\end{figure}

We propose our key hypothesis and the identification strategy using Figure \ref{fig:did} which is illustrative of the trends in the outcome of APL and BPL ration card households and the trends in PDS and market price of rice and wheat. The bold grey lines trace the daily per capita caloric intakes of the two sets of households. The dashed lines plot the movements in the market and PDS price of rice and wheat. The reference price of food for BPL households is the PDS price but for APL households it’s the market price. In reality, both sets of households are buying rice and wheat from the market but APL households primarily depend on market purchased rice and wheat (see Table \ref{tab:cons}). The price of food for the APL households is higher in 2010 in comparison to 2005 but for the BPL households it’s the same across all periods.

We are interested in showing that the APL cardholder households suffered a welfare loss due to the rise in food prices and that the PDS subsidy insures its main beneficiaries from food price increase. In terms of Figure \ref{fig:did}, this effect can be captured by the following double difference.
\begin{equation}
	\tau = (E-B)-(D-A) = (H -D)  
\end{equation}
where $(E-B)$ is the difference in the outcomes of APL and BPL households in 2005 and $(D-A)$ is the difference in their outcomes in 2010. Since the main beneficiaries of the PDS subsidy are the BPL households, their nutritional outcomes are unaffected by the price increase but we expect the outcome of the APL household to worsen in 2010. Therefore, we expect $\tau>0$. Note that, the dashed grey line segments $EH$ and $BG$ reflect the counterfactual scenarios for the APL and BPL ration card households respectively. In terms of the figure, the insurance benefit of PDS subsidy, $\tau$, is represented by $(H-D)$ which by construction in equal to $(A-G)$. The parallel trends assumption can be tested by estimating the following double difference.
\begin{equation}
	\tau^p = (F-C) - (E-B)
\end{equation}       
For parallel trends to be satisfied, $\tau^p = 0$. 

Consider the following specification which builds on the ideas presented in Figure \ref{fig:did}.
\begin{equation}\label{dido}
	\text{ln(CALORIES)}_{it} = \text{BPL}_i + \text{POST}_t + \tau \cdot \text{BPL}_i \times \text{POST}_t + X_{it}\beta + \epsilon_{it}
\end{equation}
where the dependent variable is the log of daily per capita total calories intakes. $\text{BPL}_i$ is the treatment dummy which is $1$ is the household has a BPL ration card and $0$ otherwise. $\text{POST}_t$ is $0$ for the baseline survey and $1$ for the endline survey. We are interested in the coefficient $\tau$, which quantifies the insurance role of in-kind transfers. $X$ is a vector of control variables.

Since we know that there are targeting errors in the in-kind food transfers due to misallocation of BPL ration cards, the observed BPL ration card owning status is not equal to the intended or targeted BPL ration card status. The targeting errors in ration cards can be expressed by the following relationship.
\begin{equation}
	\text{BPL}_i = \text{BPL}_{i}^* \cdot S_i
\end{equation}
where $\text{BPL}_i$ is the observed treatment dummy and $\text{BPL}_{i}^*$ is the unobserved treatment status. $S_i$ is a dummy variable which captures one-sided targeting errors. Observed BPL ration card status is defined as a product of the correctly targeted BPL ration card allocation and the targeting error captured by $S_i$.
\begin{equation}\label{pop_emp}
	\text{BPL}_i = \text{BPL}_{i}^* \cdot S_i = \mathbbm{1}\{R_i \gamma + U_i \geq 0,Z_i \alpha + V_i \geq 0\}
\end{equation}
$R$ is a vector of variables that were supposed to be used for the classification of PDS beneficiaries like incomes, consumption expenditures, land ownership, and ownership of different assets. In that sense, variables in $R$ should predict the intended beneficiaries of the PDS subsidy and hence should predict the true BPL ration card status. Vector $Z$ contains additional variables that we believe will influence mistargeting. Due to cumbersome bureaucracy and high corruption at the local level, poor households with links to local politicians or membership in local caste associations have a greater likelihood of getting BPL ration cards \citep{panda2015political}. Therefore, connections with local politicians or membership in local caste associations will independently influence the misallocation of the BPL ration cards. To implement the two step estimator proposed in this paper we first express the outcome equation in first difference form as:
\begin{equation}\label{dif}
	\Delta \text{ln(CALORIE)}_i = \alpha + \tau \text{BPL}_i + X_i\beta + \epsilon_i
\end{equation}
The estimate of $\tau$ is biased as the observed treatment dummy is misclassified. To correct for this, we first estimate the POP model expressed in equation \eqref{pop_emp}. We use the predicted BPL status from the POP in the outcome equation to get the true estimate of PDS impact. Since we assume heterogenous treatment effects, to get the true treatment effect estimates, we demean all covariates by treatment group means.

\subsection{Results}

Table \ref{tab:cons_change} presents the estimates from equation \eqref{dif} where the dependent variables are the per capita consumption of rice and wheat from different sources. We observe that high food prices induced treated households to substitute expensive market foodgrains mainly with subsidized PDS grains. Table \ref{tab:cons_change} shows that targeted PDS subsidies were effective in insulating eligible households' cereal consumption and food expenditures from high food prices.  

Table \ref{tab:did} presents the main DID results where the outcome variable is the log of per capita per day calories intake. Specification (1) presents the DID estimates without any control variables. In specifications (2), (3), and (4), we introduce household level control variables and household participation in other government programs which may be correlated with food price changes.\footnote{See Table \ref{tab:sum} for the control variables included in the DID regressions.} Considering the first three specifications, we find the estimate of the insurance impact of in-kind transfers to be positive and statistically significant across all three specifications. This implies that PDS subsidies successfully insulated the BPL ration card owning households from high food prices. These estimates are robust to addition of other household level covariates and participation and benefits received from other government welfare programs.

Although positive, the estimated impact of PDS subsidies is biased due to mistargeting. Following the two-step estimator proposed in this paper, we first estimate a partial observability probit (POP) model where we use links with local politicians and memberships in caste associations as variables independently predicting the classification errors. The estimates from the POP are presented in Appendix Table \ref{first_stg}. In the second step, we estimate the impact of the program using the predicted BPL rations card status. The estimates from the two-step estimator is presented in specification (4) of Table \ref{tab:did}. The estimate in specification (4) is much larger than what we get in a simple DID with a mistargeted treatment. This indicates that mistargeting lead to a downward bias in our estimates. This result is intuitive as some of the poor households who should have had access to the PDS subsidies didn't get the BPL ration card and hence were not insulated from the high food prices. A more policy relevant way of interpreting this result is that if PDS subsidies were correctly targeted, the insurance impact of the PDS program would have been much larger than what was actually observed.

Finally, to show that these effects were not present when food prices were not rising, we use the fact that some households were surveyed in 2004 and some in 2005 in the baseline survey. A conventional parallel trend check is not feasible in our case as a panel of households for a period before 2005 is not available. However, we can use staggered household interviews during the baseline survey to estimate a DID regression. Strictly speaking, this is not a parallel trend check as we are using a cross sectional sample and comparing different sets of households surveyed over different years, but still acts as a useful placebo test. Table \ref{pll} in the Appendix reports the results from the placebo check. The DID estimate is close to zero and is statistically insignificant in all specifications indicating no insurance benefits of the program during a period of stable food prices. 

\section{Conclusion}\label{end}
Much of the literature on difference-in-differences, both econometric and applied, assumes that the treatment receipt is measured accurately. There is sufficient evidence in the applied literature to suggest that participation in programs is prone to misclassification whether that is due to issues of misreporting by individuals or mistargeting due to errors in identifying the program beneficiaries. In this paper, we focus on identifying and estimating the ATT within a standard DID framework when the observed treatment status, $D$, wrongly classifies individuals into treatment and control groups. We show that the bias in the DID estimand is two-fold and hampers consistent estimation of the true ATT because 1) it restricts us from identifying those with $D^\ast=1$ from among those with $D=1$ and 2) differential misclassification in counterfactual trends may result in parallel trends being violated with $D$ even when we assume them to hold with the true and unobserved treatment status, $D^\ast$. 

Our main approach considers the case of one-sided misclassification and extends NDT to two-period panel and repeated cross section settings using flexible parametric regression specifications which allow for interactions between covariates, treatment, and time. We then characterize the asymptotic bias in the first-differenced and pooled OLS estimators of the ATT. We finally propose a two-step procedure which corrects for such one-sided misclassification and delivers a consistent and asymptotically normal estimator of the ATT in DID studies. We apply the proposed method to estimate the welfare impact of the Public Distribution System (PDS) of India in insulating its main beneficiaries (households below the poverty line) from commodity price risk. PDS is know to suffer from targeting errors which have largely remained unaccounted for while estimating the program effects. We find that if the PDS subsidies were correctly targeted, the insurance impact of the program would have been much larger than what was actually observed. 

\clearpage
\singlespacing\newpage {} 
\bibliographystyle{ecta.bst}
\bibliography{DID_Aug_2022.bib}

\begin{thebibliography}{69}
\newcommand{\enquote}[1]{``#1''}
\expandafter\ifx\csname natexlab\endcsname\relax\def\natexlab#1{#1}\fi

\bibitem[\protect\citeauthoryear{Abadie}{Abadie}{2005}]{abadie2005semiparametric}
\textsc{Abadie, A.} (2005): \enquote{Semiparametric difference-in-differences
  estimators,} \emph{The Review of Economic Studies}, 72, 1--19.

\bibitem[\protect\citeauthoryear{Acerenza, Ban, and K{\'e}dagni}{Acerenza
  et~al.}{2021}]{acerenza2021marginal}
\textsc{Acerenza, S., K.~Ban, and D.~K{\'e}dagni} (2021): \enquote{Marginal
  Treatment Effects with Misclassified Treatment,} \emph{arXiv preprint
  arXiv:2105.00358}.

\bibitem[\protect\citeauthoryear{Aigner}{Aigner}{1973}]{aigner1973regression}
\textsc{Aigner, D.~J.} (1973): \enquote{Regression with a binary independent
  variable subject to errors of observation,} \emph{Journal of Econometrics},
  1, 49--59.

\bibitem[\protect\citeauthoryear{Alatas, Banerjee, Hanna, Olken, and
  Tobias}{Alatas et~al.}{2012}]{alatas2012targeting}
\textsc{Alatas, V., A.~Banerjee, R.~Hanna, B.~A. Olken, and J.~Tobias} (2012):
  \enquote{Targeting the poor: evidence from a field experiment in Indonesia,}
  \emph{American Economic Review}, 102, 1206--40.

\bibitem[\protect\citeauthoryear{Balani}{Balani}{2013}]{balani2013functioning}
\textsc{Balani, S.} (2013): \enquote{Functioning of the public distribution
  system,} \emph{PRS Legislative Research, New Delhi}.

\bibitem[\protect\citeauthoryear{Battistin, De~Nadai, and Sianesi}{Battistin
  et~al.}{2014}]{battistin2014misreported}
\textsc{Battistin, E., M.~De~Nadai, and B.~Sianesi} (2014):
  \enquote{Misreported schooling, multiple measures and returns to educational
  qualifications,} \emph{Journal of Econometrics}, 181, 136--150.

\bibitem[\protect\citeauthoryear{Battistin and Sianesi}{Battistin and
  Sianesi}{2011}]{battistin2011misclassified}
\textsc{Battistin, E. and B.~Sianesi} (2011): \enquote{Misclassified treatment
  status and treatment effects: an application to returns to education in the
  United Kingdom,} \emph{Review of Economics and Statistics}, 93, 495--509.

\bibitem[\protect\citeauthoryear{Besley, Pande, and Rao}{Besley
  et~al.}{2012}]{besley2012just}
\textsc{Besley, T., R.~Pande, and V.~Rao} (2012): \enquote{Just rewards? Local
  politics and public resource allocation in South India,} \emph{The World Bank
  Economic Review}, 26, 191--216.

\bibitem[\protect\citeauthoryear{Black, Sanders, and Taylor}{Black
  et~al.}{2003}]{black2003measurement}
\textsc{Black, D., S.~Sanders, and L.~Taylor} (2003): \enquote{Measurement of
  higher education in the census and current population survey,} \emph{Journal
  of the American Statistical Association}, 98, 545--554.

\bibitem[\protect\citeauthoryear{Black, Berger, and Scott}{Black
  et~al.}{2000}]{black2000bounding}
\textsc{Black, D.~A., M.~C. Berger, and F.~A. Scott} (2000): \enquote{Bounding
  parameter estimates with nonclassical measurement error,} \emph{Journal of
  the American Statistical Association}, 95, 739--748.

\bibitem[\protect\citeauthoryear{Bollinger}{Bollinger}{1996}]{bollinger1996bounding}
\textsc{Bollinger, C.~R.} (1996): \enquote{Bounding mean regressions when a
  binary regressor is mismeasured,} \emph{Journal of Econometrics}, 73,
  387--399.

\bibitem[\protect\citeauthoryear{Botosaru and Gutierrez}{Botosaru and
  Gutierrez}{2018}]{botosaru2018difference}
\textsc{Botosaru, I. and F.~H. Gutierrez} (2018):
  \enquote{Difference-in-differences when the treatment status is observed in
  only one period,} \emph{Journal of Applied Econometrics}, 33, 73--90.

\bibitem[\protect\citeauthoryear{Bound, Brown, and Mathiowetz}{Bound
  et~al.}{2001}]{bound2001measurement}
\textsc{Bound, J., C.~Brown, and N.~Mathiowetz} (2001): \enquote{Measurement
  error in survey data,} in \emph{Handbook of Econometrics}, Elsevier, vol.~5,
  3705--3843.

\bibitem[\protect\citeauthoryear{Bruckmeier, Riphahn, and Wiemers}{Bruckmeier
  et~al.}{2021}]{bruckmeier2021misreporting}
\textsc{Bruckmeier, K., R.~T. Riphahn, and J.~Wiemers} (2021):
  \enquote{Misreporting of program take-up in survey data and its consequences
  for measuring non-take-up: new evidence from linked administrative and survey
  data,} \emph{Empirical Economics}, 61, 1567--1616.

\bibitem[\protect\citeauthoryear{Buchmueller, DiNardo, and
  Valletta}{Buchmueller et~al.}{2011}]{buchmueller2011effect}
\textsc{Buchmueller, T.~C., J.~DiNardo, and R.~G. Valletta} (2011):
  \enquote{The effect of an employer health insurance mandate on health
  insurance coverage and the demand for labor: Evidence from Hawaii,}
  \emph{American Economic Journal: Economic Policy}, 3, 25--51.

\bibitem[\protect\citeauthoryear{Byker and Gutierrez}{Byker and
  Gutierrez}{2016}]{byker2016treatment}
\textsc{Byker, T. and I.~A. Gutierrez} (2016): \enquote{Treatment Effects Using
  Inverse Probability Weighting and Contaminated Treatment Data: An Application
  to the Evaluation of a Government Female Sterilization Campaign in Peru,} .

\bibitem[\protect\citeauthoryear{Callaway and Sant’Anna}{Callaway and
  Sant’Anna}{2020}]{callaway2020difference}
\textsc{Callaway, B. and P.~H. Sant’Anna} (2020):
  \enquote{Difference-in-differences with multiple time periods,} \emph{Journal
  of Econometrics}.

\bibitem[\protect\citeauthoryear{Calvi, Lewbel, and Tommasi}{Calvi
  et~al.}{2021}]{calvietal2021}
\textsc{Calvi, R., A.~Lewbel, and D.~Tommasi} (2021): \enquote{LATE with
  Missing or Mismeasured Treatment,} \emph{Journal of Business \& Economic
  Statistics}, 0, 1--45.

\bibitem[\protect\citeauthoryear{Cameron and Shah}{Cameron and
  Shah}{2014}]{cameron2014can}
\textsc{Cameron, L. and M.~Shah} (2014): \enquote{Can mistargeting destroy
  social capital and stimulate crime? Evidence from a cash transfer program in
  Indonesia,} \emph{Economic Development and Cultural Change}, 62, 381--415.

\bibitem[\protect\citeauthoryear{Card}{Card}{1996}]{card1996effect}
\textsc{Card, D.} (1996): \enquote{The effect of unions on the structure of
  wages: A longitudinal analysis,} \emph{Econometrica}, 957--979.

\bibitem[\protect\citeauthoryear{Coady, Grosh, and Hoddinott}{Coady
  et~al.}{2004}]{coady2004targeting}
\textsc{Coady, D., M.~E. Grosh, and J.~Hoddinott} (2004): \enquote{Targeting of
  transfers in developing countries: Review of lessons and experience,} .

\bibitem[\protect\citeauthoryear{Cornia and Stewart}{Cornia and
  Stewart}{1993}]{cornia1993two}
\textsc{Cornia, G.~A. and F.~Stewart} (1993): \enquote{Two errors of
  targeting,} \emph{Journal of International Development}, 5, 459--496.

\bibitem[\protect\citeauthoryear{De~Chaisemartin and
  d'Haultfoeuille}{De~Chaisemartin and d'Haultfoeuille}{2020}]{de2020two}
\textsc{De~Chaisemartin, C. and X.~d'Haultfoeuille} (2020): \enquote{Two-way
  fixed effects estimators with heterogeneous treatment effects,}
  \emph{American Economic Review}, 110, 2964--96.

\bibitem[\protect\citeauthoryear{De~Chaisemartin and
  d’Haultfoeuille}{De~Chaisemartin and d’Haultfoeuille}{2018}]{de2018fuzzy}
\textsc{De~Chaisemartin, C. and X.~d’Haultfoeuille} (2018): \enquote{Fuzzy
  differences-in-differences,} \emph{The Review of Economic Studies}, 85,
  999--1028.

\bibitem[\protect\citeauthoryear{Desai and Vanneman}{Desai and
  Vanneman}{2010}]{desai2010national}
\textsc{Desai, S. and R.~Vanneman} (2010): \enquote{National {C}ouncil of
  {A}pplied {E}conomic {R}esearch, {N}ew {D}elhi. {I}ndia {H}uman {D}evelopment
  {S}urvey ({IHDS}), 2005.} \emph{Ann Arbor, MI: Inter-university Consortium
  for Political and Social Research}.

\bibitem[\protect\citeauthoryear{Desai and Vanneman}{Desai and
  Vanneman}{2018}]{desai2018national}
---\hspace{-.1pt}---\hspace{-.1pt}--- (2018): \enquote{National {C}ouncil of
  {A}pplied {E}conomic {R}esearch, {N}ew {D}elhi. {I}ndia {H}uman {D}evelopment
  {S}urvey ({IHDS}), 2012,} \emph{Ann Arbor, MI: Inter-university Consortium
  for Political and Social Research, University of Michigan}.

\bibitem[\protect\citeauthoryear{DiTraglia and Garcia-Jimeno}{DiTraglia and
  Garcia-Jimeno}{2019}]{ditraglia2019identifying}
\textsc{DiTraglia, F.~J. and C.~Garcia-Jimeno} (2019): \enquote{Identifying the
  effect of a mis-classified, binary, endogenous regressor,} \emph{Journal of
  Econometrics}, 209, 376--390.

\bibitem[\protect\citeauthoryear{Dutta and Ramaswami}{Dutta and
  Ramaswami}{2001}]{dutta2001targeting}
\textsc{Dutta, B. and B.~Ramaswami} (2001): \enquote{Targeting and efficiency
  in the public distribution system: Case of Andhra Pradesh and Maharashtra,}
  \emph{Economic and Political Weekly}, 1524--1532.

\bibitem[\protect\citeauthoryear{Emran, Robano, and Smith}{Emran
  et~al.}{2014}]{emran2014assessing}
\textsc{Emran, M.~S., V.~Robano, and S.~C. Smith} (2014): \enquote{Assessing
  the frontiers of ultrapoverty reduction: evidence from challenging the
  frontiers of poverty reduction/targeting the ultra-poor, an innovative
  program in Bangladesh,} \emph{Economic Development and Cultural Change}, 62,
  339--380.

\bibitem[\protect\citeauthoryear{Frazis and Loewenstein}{Frazis and
  Loewenstein}{2003}]{frazis2003estimating}
\textsc{Frazis, H. and M.~A. Loewenstein} (2003): \enquote{Estimating linear
  regressions with mismeasured, possibly endogenous, binary explanatory
  variables,} \emph{Journal of Econometrics}, 117, 151--178.

\bibitem[\protect\citeauthoryear{Gadenne, Norris, Singhal, and
  Sukhtankar}{Gadenne et~al.}{2021}]{gadenne2021kind}
\textsc{Gadenne, L., S.~Norris, M.~Singhal, and S.~Sukhtankar} (2021):
  \enquote{In-kind transfers as insurance,} Tech. rep., National Bureau of
  Economic Research.

\bibitem[\protect\citeauthoryear{Goodman-Bacon}{Goodman-Bacon}{2018}]{goodman2018difference}
\textsc{Goodman-Bacon, A.} (2018): \enquote{Difference-in-differences with
  variation in treatment timing,} Tech. rep., National Bureau of Economic
  Research.

\bibitem[\protect\citeauthoryear{Groen and Polivka}{Groen and
  Polivka}{2008}]{groen2008effect}
\textsc{Groen, J.~A. and A.~E. Polivka} (2008): \enquote{The effect of
  Hurricane Katrina on the labor market outcomes of evacuees,} \emph{American
  Economic Review}, 98, 43--48.

\bibitem[\protect\citeauthoryear{Haider and Stephens~Jr}{Haider and
  Stephens~Jr}{2020}]{haider2020correcting}
\textsc{Haider, S.~J. and M.~Stephens~Jr} (2020): \enquote{Correcting for
  Misclassified Binary Regressors Using Instrumental Variables,} Tech. rep.,
  National Bureau of Economic Research.

\bibitem[\protect\citeauthoryear{Hirway}{Hirway}{2003}]{hirway2003identification}
\textsc{Hirway, I.} (2003): \enquote{Identification of BPL households for
  poverty alleviation programmes,} \emph{Economic and Political Weekly},
  4803--4808.

\bibitem[\protect\citeauthoryear{Hong}{Hong}{2013}]{hong2013measuring}
\textsc{Hong, S.-H.} (2013): \enquote{Measuring the effect of napster on
  recorded music sales: difference-in-differences estimates under compositional
  changes,} \emph{Journal of Applied Econometrics}, 28, 297--324.

\bibitem[\protect\citeauthoryear{Hu}{Hu}{2008}]{hu2008identification}
\textsc{Hu, Y.} (2008): \enquote{Identification and estimation of nonlinear
  models with misclassification error using instrumental variables: A general
  solution,} \emph{Journal of Econometrics}, 144, 27--61.

\bibitem[\protect\citeauthoryear{Jha, Kotwal, and Ramaswami}{Jha
  et~al.}{2013}]{jha2013safety}
\textsc{Jha, S., A.~Kotwal, and B.~Ramaswami} (2013): \enquote{Safety Nets and
  Food Programs in Asia: A Comparative Perspective,} \emph{Asian Development
  Bank Economics Working Paper Series}.

\bibitem[\protect\citeauthoryear{Kane, Rouse, and Staiger}{Kane
  et~al.}{1999}]{kane1999estimating}
\textsc{Kane, T.~J., C.~E. Rouse, and D.~O. Staiger} (1999):
  \enquote{Estimating returns to schooling when schooling is misreported,} .

\bibitem[\protect\citeauthoryear{Kaushal and Muchomba}{Kaushal and
  Muchomba}{2015}]{kaushal2015consumer}
\textsc{Kaushal, N. and F.~M. Muchomba} (2015): \enquote{How consumer price
  subsidies affect nutrition,} \emph{World Development}, 74, 25--42.

\bibitem[\protect\citeauthoryear{Khera}{Khera}{2008}]{khera2008access}
\textsc{Khera, R.} (2008): \enquote{Access to the targeted public distribution
  system: a case study in Rajasthan,} \emph{Economic and Political Weekly},
  51--56.

\bibitem[\protect\citeauthoryear{Kochar}{Kochar}{2005}]{kochar2005can}
\textsc{Kochar, A.} (2005): \enquote{Can targeted food programs improve
  nutrition? An empirical analysis of India’s public distribution system,}
  \emph{Economic development and cultural change}, 54, 203--235.

\bibitem[\protect\citeauthoryear{Kreider}{Kreider}{2010}]{kreider2010regression}
\textsc{Kreider, B.} (2010): \enquote{Regression coefficient identification
  decay in the presence of infrequent classification errors,} \emph{The Review
  of Economics and Statistics}, 92, 1017--1023.

\bibitem[\protect\citeauthoryear{Kreider, Pepper, Gundersen, and
  Jolliffe}{Kreider et~al.}{2012}]{kreider2012identifying}
\textsc{Kreider, B., J.~V. Pepper, C.~Gundersen, and D.~Jolliffe} (2012):
  \enquote{Identifying the effects of SNAP (food stamps) on child health
  outcomes when participation is endogenous and misreported,} \emph{Journal of
  the American Statistical Association}, 107, 958--975.

\bibitem[\protect\citeauthoryear{Lewbel}{Lewbel}{2007}]{lewbel2007estimation}
\textsc{Lewbel, A.} (2007): \enquote{Estimation of average treatment effects
  with misclassification,} \emph{Econometrica}, 75, 537--551.

\bibitem[\protect\citeauthoryear{Loewenstein and Spletzer}{Loewenstein and
  Spletzer}{1997}]{loewenstein1997delayed}
\textsc{Loewenstein, M.~A. and J.~R. Spletzer} (1997): \enquote{Delayed formal
  on-the-job training,} \emph{ILR Review}, 51, 82--99.

\bibitem[\protect\citeauthoryear{Mahajan}{Mahajan}{2006}]{mahajan2006identification}
\textsc{Mahajan, A.} (2006): \enquote{Identification and estimation of
  regression models with misclassification,} \emph{Econometrica}, 74, 631--665.

\bibitem[\protect\citeauthoryear{Martinelli and Parker}{Martinelli and
  Parker}{2009}]{martinelli2009deception}
\textsc{Martinelli, C. and S.~W. Parker} (2009): \enquote{Deception and
  misreporting in a social program,} \emph{Journal of the European Economic
  Association}, 7, 886--908.

\bibitem[\protect\citeauthoryear{Meyer and Mittag}{Meyer and
  Mittag}{2017}]{meyer2017misclassification}
\textsc{Meyer, B.~D. and N.~Mittag} (2017): \enquote{Misclassification in
  binary choice models,} \emph{Journal of Econometrics}, 200, 295--311.

\bibitem[\protect\citeauthoryear{Meyer, Mok, and Sullivan}{Meyer
  et~al.}{2015}]{meyer2015household}
\textsc{Meyer, B.~D., W.~K. Mok, and J.~X. Sullivan} (2015): \enquote{Household
  surveys in crisis,} \emph{Journal of Economic Perspectives}, 29, 199--226.

\bibitem[\protect\citeauthoryear{Millimet}{Millimet}{2011}]{millimet2011elephant}
\textsc{Millimet, D.~L.} (2011): \enquote{The elephant in the corner: a
  cautionary tale about measurement error in treatment effects models,} in
  \emph{Missing data methods: Cross-sectional methods and applications},
  Emerald Group Publishing Limited.

\bibitem[\protect\citeauthoryear{Molinari}{Molinari}{2008}]{molinari2008partial}
\textsc{Molinari, F.} (2008): \enquote{Partial identification of probability
  distributions with misclassified data,} \emph{Journal of Econometrics}, 144,
  81--117.

\bibitem[\protect\citeauthoryear{Negi}{Negi}{2022}]{negi2022global}
\textsc{Negi, D.} (2022): \enquote{Global food price surge, in-kind transfers,
  and household welfare: Evidence from India,} \emph{World Development}.

\bibitem[\protect\citeauthoryear{Nguimkeu, Denteh, and Tchernis}{Nguimkeu
  et~al.}{2019}]{nguimkeu2019estimation}
\textsc{Nguimkeu, P., A.~Denteh, and R.~Tchernis} (2019): \enquote{On the
  estimation of treatment effects with endogenous misreporting,} \emph{Journal
  of Econometrics}, 208, 487--506.

\bibitem[\protect\citeauthoryear{Panda}{Panda}{2015}]{panda2015political}
\textsc{Panda, S.} (2015): \enquote{Political connections and elite capture in
  a poverty alleviation programme in India,} \emph{The Journal of Development
  Studies}, 51, 50--65.

\bibitem[\protect\citeauthoryear{Pande}{Pande}{2007}]{pande2007understanding}
\textsc{Pande, R.} (2007): \enquote{Understanding political corruption in low
  income countries,} \emph{Handbook of development economics}, 4, 3155--3184.

\bibitem[\protect\citeauthoryear{Pingali, Aiyar, Abraham, and Rahman}{Pingali
  et~al.}{2019}]{pingali2019reimagining}
\textsc{Pingali, P., A.~Aiyar, M.~Abraham, and A.~Rahman} (2019):
  \enquote{Reimagining safety net programs,} in \emph{Transforming food systems
  for a rising India}, Springer, 135--164.

\bibitem[\protect\citeauthoryear{Poirier}{Poirier}{1980}]{poirier1980partial}
\textsc{Poirier, D.~J.} (1980): \enquote{Partial observability in bivariate
  probit models,} \emph{Journal of Econometrics}, 12, 209--217.

\bibitem[\protect\citeauthoryear{Ram, Mohanty, and Ram}{Ram
  et~al.}{2009}]{ram2009understanding}
\textsc{Ram, F., S.~Mohanty, and U.~Ram} (2009): \enquote{Understanding the
  distribution of BPL cards: all-India and selected states,} \emph{Economic and
  Political Weekly}, 66--71.

\bibitem[\protect\citeauthoryear{Rambachan and Roth}{Rambachan and
  Roth}{2022}]{rambachan2022more}
\textsc{Rambachan, A. and J.~Roth} (2022): \enquote{A More Credible Approach to
  Parallel Trends,} Tech. rep., Working Paper.

\bibitem[\protect\citeauthoryear{Sant’Anna and Zhao}{Sant’Anna and
  Zhao}{2020}]{sant2020doubly}
\textsc{Sant’Anna, P.~H. and J.~Zhao} (2020): \enquote{Doubly robust
  difference-in-differences estimators,} \emph{Journal of Econometrics}, 219,
  101--122.

\bibitem[\protect\citeauthoryear{Sun and Abraham}{Sun and
  Abraham}{2020}]{sun2020estimating}
\textsc{Sun, L. and S.~Abraham} (2020): \enquote{Estimating dynamic treatment
  effects in event studies with heterogeneous treatment effects,} \emph{Journal
  of Econometrics}.

\bibitem[\protect\citeauthoryear{Swaminathan and Misra}{Swaminathan and
  Misra}{2001}]{swaminathan2001errors}
\textsc{Swaminathan, M. and N.~Misra} (2001): \enquote{Errors of targeting:
  Public distribution of food in a Maharashtra village, 1995-2000,}
  \emph{Economic and Political Weekly}, 2447--2454.

\bibitem[\protect\citeauthoryear{Tallis}{Tallis}{1961}]{tallis1961moment}
\textsc{Tallis, G.~M.} (1961): \enquote{The moment generating function of the
  truncated multi-normal distribution,} \emph{Journal of the Royal Statistical
  Society: Series B (Methodological)}, 23, 223--229.

\bibitem[\protect\citeauthoryear{Tohari, Parsons, and Rammohan}{Tohari
  et~al.}{2019}]{tohari2019targeting}
\textsc{Tohari, A., C.~Parsons, and A.~Rammohan} (2019): \enquote{Targeting
  poverty under complementarities: Evidence from Indonesia's unified targeting
  system,} \emph{Journal of Development Economics}, 140, 127--144.

\bibitem[\protect\citeauthoryear{Tommasi and Zhang}{Tommasi and
  Zhang}{2020}]{tommasi2020bounding}
\textsc{Tommasi, D. and L.~Zhang} (2020): \enquote{Bounding Program Benefits
  When Participation Is Misreported,} .

\bibitem[\protect\citeauthoryear{Ura}{Ura}{2018}]{ura2018heterogeneous}
\textsc{Ura, T.} (2018): \enquote{Heterogeneous treatment effects with
  mismeasured endogenous treatment,} \emph{Quantitative Economics}, 9,
  1335--1370.

\bibitem[\protect\citeauthoryear{Wooldridge}{Wooldridge}{2021}]{wooldridgetwfe}
\textsc{Wooldridge, M.~J.} (2021): \enquote{Two-Way Fixed Effects, the Two-Way
  Mundlak Regression, and Difference-in-Differences Estimators,} .

\bibitem[\protect\citeauthoryear{Yanagi}{Yanagi}{2019}]{yanagi2019inference}
\textsc{Yanagi, T.} (2019): \enquote{Inference on local average treatment
  effects for misclassified treatment,} \emph{Econometric Reviews}, 38,
  938--960.

\end{thebibliography}

\clearpage
\section*{Figures}\label{figures}

\clearpage
\section*{Tables}\label{tables}
	\begin{table}[H]
	\centering
	\footnotesize{
	\begin{threeparttable}
	\caption{\textbf{Summary Statistics}}\label{tab:sum}
    \begin{tabular}{lrrrr}
    \toprule\toprule
	Variables & \multicolumn{ 2}{c}{Baseline (2005)} & \multicolumn{ 2}{c}{Endline (2012)} \\
	\cmidrule{2-5}
	& Control & Treated & Control & Treated \\
	\midrule
	No ration card &        495 &          - &        495 &          - \\
	APL (above poverty line) ration card &      5,850 &          - &      5,850 &          - \\
	BPL (below poverty line) ration card &          - &      2,888 &          - &      2,888 \\
	AAY (antyodaya ann yojna) ration card &          - &        147 &          - &        147 \\
	Households &      6,345 &      3,035 &      6,345 &      3,035 \\
	\textit{Outcome} \\
	Calories consumed per person per day &    2519.83 &    2137.82 &    2398.78 &    2129.91 \\
	\textit{Covariates} \\
	Operated land (ha) &       2.04 &       0.99 &       1.36 &       0.82 \\
	Number of family members &       6.92 &       5.87 &       5.40 &       4.97 \\
	Any adult literate &       0.87 &       0.72 &       0.88 &       0.75 \\
	Monthly cons. expenditure (rs/person) &    1761.93 &    1079.49 &    2336.43 &    1521.29 \\
	Assets: owns any vehicle &       0.70 &       0.57 &       0.70 &       0.60 \\
	Assets: owns motor vehicle &       0.24 &       0.03 &       0.37 &       0.13 \\
	Assets: owns cooler or AC &       0.14 &       0.01 &       0.21 &       0.04 \\
	Assets: owns TV &       0.58 &       0.25 &       0.65 &       0.47 \\
	Assets: own electric fan &       0.66 &       0.32 &       0.77 &       0.54 \\
	Assets: own refrigerator &       0.17 &       0.01 &       0.31 &       0.04 \\
	Permanent house structure &       0.66 &       0.36 &       0.75 &       0.46 \\
	Household has electricity &       0.76 &       0.62 &       0.86 &       0.80 \\
	Household owns farm equipment &       0.13 &       0.01 &       0.11 &       0.01 \\
	Any member in MGNREGA work &       0.00 &       0.02 &       0.15 &       0.26 \\
	Benefits from govt. programs (000 rupees) &       0.01 &       0.01 &       0.02 &       0.04 \\
	Religion: hindu &       0.83 &       0.90 &       - &       - \\
	Religion: muslim &       0.10 &       0.05 &       - &       - \\
	Religion: others &       0.05 &       0.01 &       - &       - \\
	Caste: scheduled caste and scheduled tribes &       0.17 &       0.41 &       - &      - \\
	Caste: other backward caste &       0.35 &       0.41 &       - &       - \\
	Caste: general &       0.32 &       0.12 &       - &       - \\
	Proportion of area under rice-wheat &       0.46 &       0.45 &       - &       - \\
	Proportion in rural areas &       0.95 &       0.98 &       - &       - \\
	\textit{Missclassificaiton predictors} \\
	Household member(s) in local caste association &       0.10 &       0.17 &       - &      - \\
	Household member(s) links to local officials &       0.12 &       0.14 &       - &       - \\
	\bottomrule\bottomrule
\end{tabular}  
	\begin{tablenotes}[flushleft]
	\item \footnotesize Notes: MGNREGA stands for the Mahatma Gandhi Rural Employment Guarantee Scheme. MGNREGA is India's large-scale anti-poverty rural workfare program. It was introduced in 2005 and provides 100 days per year of voluntary employment at minimum wages to individuals in the working age group. The MGNREGA is mostly operational in rural areas and provides unskilled labor employment on local public work projects. Benefit from government programs is calculated as the total transfers received from scholarships, old age pension, maternity scheme, disability scheme, income generation programs other than MGNREGA, assistance from drought/flood compensation, and insurance payouts.  AAY is given to poorest families from within the below poverty line households. BPL and AAY ration card household are the main benificiaries of the PDS subsidies hence form the treatment group.
\end{tablenotes}
\end{threeparttable}
}
\end{table}	

\begin{landscape}
		\begin{table}[H]
		\centering
		\begin{threeparttable}
			\caption{\textbf{Monthly Rice and Wheat Consumption for Treatment and Conrol Households (in kilograms per person)}}\label{tab:cons}
			\begin{tabular}{lrrrrrrrr}
				\toprule\toprule
				Ration card type & Total & Homegrown & Market & PDS & Total & Homegrown & Market & PDS \\
				\midrule
				&              \multicolumn{ 4}{c}{Baseline (2005)} &               \multicolumn{ 4}{c}{Endline (2012)} \\
				Control: No ration card &       12.8 &        0.6 &       12.0 &        0.0 &       12.6 &        0.8 &       11.6 &        0.0 \\
				Control: APL (above poverty line) ration card &       11.0 &        0.6 &        9.8 &        0.7 &       11.6 &        0.8 &        9.5 &        1.3 \\
				Treatment: BPL (below poverty line) ration card &       11.2 &        0.7 &        7.5 &        3.0 &       12.0 &        1.0 &        5.6 &        5.4 \\
				Treatment: AAY (antyodaya ann yojna) ration card &       14.2 &        0.8 &        7.6 &        5.7 &       14.6 &        1.0 &        5.4 &        8.1 \\
				\bottomrule\bottomrule
			\end{tabular}  
			\begin{tablenotes}[flushleft]
				\item \footnotesize Notes: No ration card and APL ration card households are in the control group. AAY is given to poorest families from within the below poverty line households. BPL and AAY ration card household are the main benificiaries of the PDS subsidies hence form the treatment group.
			\end{tablenotes}
		\end{threeparttable}
	\end{table}	
\end{landscape}

		\begin{table}[H]
			\centering
			\begin{threeparttable}
			\caption{\textbf{Change in Monthly Food Consumption of BPL Ration Card Owning Households}}\label{tab:cons_change}
			\begin{tabular}{lcccccc}
				\toprule\toprule
				& (1)   & (2)   & (3)   & (4)   & (5)   & (6) \\
				\midrule
			 	& \multicolumn{4}{c}{Rice and wheat (kg/person)}   &  & Food expenditure\\
				& Total & Homegrown & Market & PDS   & Cereals & rs/person \\
				$BPL$ &      0.233 &   0.053*** &  -1.603*** &   1.783*** &     0.391* &   -21.925* \\
				&    (0.176) &    (0.017) &    (0.216) &    (0.153) &    (0.204) &   (12.891) \\
				\\
				$N$ &       9380 &       9380 &       9380 &       9380 &       9380 &       9380 \\
				\bottomrule\bottomrule
			\end{tabular}%
				    \begin{tablenotes}[flushleft]
					\item \footnotesize Notes: Table presents estimates from differenced equation. Dependent variables are in terms of changes in the baseline and endline period. $BPL$ is a dummy variable which is 1 if the household owns a below poverty line ration card, 0 otherwise. Standard errors in parentheses clustered at the village level. ***, **, and * indicate statistical significance at the 1\%, 5\%, and 10\% levels, respectively.
				\end{tablenotes}
				\end{threeparttable}
		\end{table}	

\begin{table}[H]
	\centering
	\begin{threeparttable}
		\caption{\textbf{Difference-in-Difference Estimates}}\label{tab:did}
		\begin{tabular}{lcccc}
			\toprule\toprule
			& (1)   & (2)   & (3)   & (4) \\
			\midrule
			& \multicolumn{3}{c}{Observed treatment status} & Predicted treatment status \\
			\multicolumn{5}{l}{Dependent variable: Log change in calories consumer per person per day} \\
			$BPL$ &   0.042*** &   0.034*** &   0.034*** &   0.231** \\
			&    (0.016) &    (0.013) &    (0.013) &    (0.099) \\
			\\
			Controls & No    & Yes   & Yes   & Yes \\
			Government programs & No    & No    & Yes   & Yes \\
			\\
         	$N$ &   9380 &   9380 &   9380 &   9380 \\
		 	$R^2$ &      0.002 &      0.335 &      0.335 &      0.336 \\
		 	$F$ &      7.082 &    156.231 &    142.458 &    141.984 \\
			\bottomrule\bottomrule
		\end{tabular}%
		\begin{tablenotes}[flushleft]
			\item \footnotesize Notes: $BPL$  is a dummy variable which is 1 if the household owns a below poverty line ration card, 0 otherwise. Household controls include log operated land, log consumption expenditure, dummy for a literate member present in the household, caste dummies, number of family members, rural dummy, asset count, dummies for ownership of vehicle, motor car, moterbike, cooler or AC, refrigerator, TV, electric fan, telephone, electricity and finally a dummy for whether the household have a makeshift dwelling or a \textit{kaccha} house. Government programs include participation in the Mahatma Gandhi National Rural Guarantee Scheme (MGNREGS) or benefits received from other welfare programs like health insurance, scholarships, old age pension, maternity scheme, disability scheme, income generation programs other than MGNREGS, assistance from drought/flood compensation, and insurance payouts. All covariates demeaned by treatment group means. Standard errors in parentheses for specifications (1), (2) and (3) are clustered at the village level. Specification (4) has bootstrapped standard errors with 200 replications. ***, **, and * indicate statistical significance at the 1\%, 5\%, and 10\% levels, respectively.
		\end{tablenotes}
	\end{threeparttable}
\end{table}

\appendix
\singlespacing\newpage {} 
\numberwithin{equation}{section}
\numberwithin{assumption}{section}
\numberwithin{table}{section}
\numberwithin{figure}{section}

\newpage
\appendix
\section*{Appendix}
\section{Proofs}
\begin{proof}[Proof of Proposition 1] We consider the case of the two period panel and the repeated cross section in turn. For ease of derivations, define $q_1 \equiv \mathbb{P}(D^\ast=1|D=1)$ and $q_0 \equiv \mathbb{P}(D^\ast=0|D=0)$. Then, 
	\paragraph{Two-period panel} Using bayes rule, for each $d, d^\prime=0,1$, we can express
	{\small\begin{equation*}
			\begin{split}
				\mathbb{E}[G_t(0)|D=1] &= \mathbb{E}\{\mathbb{E}(G_t(0)|D=1, D^\ast=d^\prime)|D=1\} \\
				& = \mathbb{E}(G_t(0)|D=1, D^\ast=1)\cdot q_1
				+ \mathbb{E}(G_t(0)|D=1, D^\ast=0)\cdot(1-q_1)
			\end{split}
	\end{equation*}} Similarly, we can express $\mathbb{E}[G_1(0)|D=0]$ in terms of $D^\ast$ and $q_0$. Then, 
	{\small\begin{equation*}\label{pt_true}
			\begin{split}
				\mathrm{DT}(D) & = \mathbb{E}[G_1(0)|D=1] -\mathbb{E}[G_1(0)|D=0]\\
				& = \bigg\{\mathbb{E}[G_1(0)|D=1, D^\ast=1]\cdot q_1+\mathbb{E}[G_1(0)|D=1, D^\ast=0]\cdot (1-q_1)\bigg\} - \bigg\{\mathbb{E}[G_1(0)|D=0, D^\ast=1]\cdot \\ &\ \ \ \ \ (1-q_0)+\mathbb{E}[G_1(0)|D=0, D^\ast=0]\cdot q_0\bigg\} \\
				& = \big\{\mathbb{E}[G_1(0)|D=1, D^\ast=1]-\mathbb{E}[G_1(0)|D=1, D^\ast=0]\big\}\cdot q_1+\big\{\mathbb{E}[G_1(0)|D=0, D^\ast=1] \\
				& \ \ \ \ -\mathbb{E}[G_1(0)|D=0, D^\ast=0]\big\}\cdot q_0+\left\{\mathbb{E}[G_1(0)|D=1, D^\ast=0]-\mathbb{E}[G_1(0)|D=0, D^\ast=1]\right\}
			\end{split}
	\end{equation*}} Now if we impose the assumption that $\mathbb{E}[G_t(0)|D, D^\ast] = \mathbb{E}[G_t(0)|D^\ast] $ then we obtain 
	{\small\begin{align*}\label{att_true}
			\mathrm{DT}(D) &=  \big\{\mathbb{E}[G_1(0)|D^\ast=1]-\mathbb{E}[G_1(0)|D^\ast=0]\big\}\cdot q_1+\big\{\mathbb{E}[G_1(0)|D^\ast=1]-\mathbb{E}[G_1(0)|D^\ast=0]\big\}\cdot q_0\\
			& +\left\{\mathbb{E}[G_1(0)|D^\ast=0]-\mathbb{E}[G_1(0)|D^\ast=1]\right\} \\
			& = \mathrm{DT}(D^\ast)\cdot (q_1+q_0-1)
	\end{align*}} 
	\paragraph{Repeated cross section} In this case, we can write
	{\small \begin{align*}
			&\mathbb{E}[Y(0)|D=d, D^\ast=d^\prime, T=t] = \mathbb{E}\{\mathbb{E}[Y(0)|D=d, D^\ast=d^\prime, T=t]|D=d, T=t\} \\
			& = \mathbb{E}[Y(0)|D=d, D^\ast=1, T=t]\cdot \mathbb{P}(D^\ast=1|D=d, T=t)+\mathbb{E}[Y(0)|D=d, D^\ast=0, T=t] \\
			& \ \ \cdot \mathbb{P}(D^\ast=0|D=d, T=t) \\
			& = \mathbb{E}[Y(0)|D=d, D^\ast=1, T=t]\cdot \mathbb{P}(D^\ast=1|D=d)+\mathbb{E}[Y(0)|D=d, D^\ast=0, T=t] \\
			& \ \ \cdot \mathbb{P}(D^\ast=0|D=d) 
	\end{align*}}
	Then, 
	{\small \begin{align*}
			\mathrm{DT}(D) &= \mathbb{E}[Y(0)|D=1, T=1]-\mathbb{E}[Y(0)|D=1, T=0]-\mathbb{E}[Y(0|D=0, T=1)+\mathbb{E}[Y(0)|D=0, T=0]] \\
			& = \bigg\{\mathbb{E}[Y(0)|D=1, D^\ast=1, T=1]-\mathbb{E}[Y(0)|D=1, D^\ast=1, T=0]-\mathbb{E}[Y(0)|D=1, D^\ast=0, T=1]\\
			&+\mathbb{E}[Y(0)|D=1, D^\ast=0, T=0]\bigg\}q_1-\bigg\{\mathbb{E}[Y(0)|D=0, D^\ast=0, T=1]\\
			&-\mathbb{E}[Y(0)|D=0, D^\ast=0, T=0] -  \mathbb{E}[Y(0)|D=0, D^\ast=1, T=1]+\mathbb{E}[Y(0)|D=0, D^\ast=1,T=0]
			\bigg\}q_0 \\
			&+\bigg\{\mathbb{E}[Y(0)|D=1, D^\ast=0, T=1]-\mathbb{E}[Y(0)|D=1, D^\ast=0, T=0] -\mathbb{E}[Y(0)|D=0, D^\ast=1, T=1]\\
			&+\mathbb{E}[Y(0)|D=0, D^\ast=1, T=0]\bigg\}
	\end{align*}} 
	Since $\mathbb{E}[G_1(0)|D, D^\ast] = \mathbb{E}[G_1(0)|D^\ast]$ in the case of a repeated cross section implies that $\mathbb{E}[Y(0)|D, D^\ast, T=1]-\mathbb{E}[Y(0)|D, D^\ast, T=0] = \mathbb{E}[Y(0)|D^\ast, T=1]-\mathbb{E}[Y(0)|D^\ast, T=0]$. Therefore once we plug this in the above expression, we obtain the desired result.
\end{proof}

\begin{proof}(Bias of $\hat{\theta}_{FD}$)
	In matrix notation, we can write the regression in (\ref{fd_ifreg}) as,
	\begin{equation*}
		\Delta Y = \ddot{R}\delta+\ddot{W}\theta+\Delta\epsilon
	\end{equation*} where $\Delta Y$, $\ddot{R}$, $\ddot{W}$, and $\Delta\epsilon$ are the $N$-vector data matrices. Then let $\ddot{M} = I-\ddot{R}(\ddot{R}^\prime \ddot{R})^{-1}\ddot{R}^\prime$ be the residual making matrix of $\ddot{R}$.
	Consider the following regression,
	\begin{equation}\label{res_reg}
		\ddot{M}\Delta Y = \ddot{M}\ddot{W}\theta+\ddot{M}\Delta\epsilon
	\end{equation}
	Then, 
	\begin{equation*}
		\begin{split}
			\hat{\theta}_{FD} &= \left[(\ddot{M}\ddot{W})^\prime (\ddot{M}\ddot{W})\right]^{-1}(\ddot{M}\ddot{W})^\prime (\ddot{M} \Delta Y) \\
			&= \left(\ddot{W}^\prime\ddot{M}\ddot{W}\right)^{-1}\ddot{W}^\prime \ddot{M}\Delta Y \\
			& = \theta + \left(\ddot{W}^\prime\ddot{M}\ddot{W}\right)^{-1}\ddot{W}^\prime\ddot{M}\Delta\epsilon
		\end{split}
	\end{equation*}
	This implies that 
	{\small \begin{align}\label{ols_bias}
			\hat{\theta}_{FD}-\theta &=  \left(\ddot{W}^\prime\ddot{M}\ddot{W}\right)^{-1}\ddot{W}^\prime\ddot{M}\Delta\epsilon \nonumber \\
			&  = \left(\frac{\ddot{W}^\prime\ddot{M}\ddot{W}}{N}\right)^{-1}\frac{\ddot{W}^\prime\ddot{M}\Delta\epsilon}{N} \nonumber \\
			& = \left(\frac{\ddot{W}^\prime\ddot{M}\ddot{W}}{N}\right)^{-1}\frac{\ddot{W}^\prime\ddot{M}}{N}\left\{(\mathring{R}-\ddot{R})\delta+(W^\ast-\ddot{W})\theta+\Delta\xi\right\} \nonumber \\
			&  =\underbrace{\left(\frac{\ddot{W}^\prime\ddot{M}\ddot{W}}{N}\right)^{-1}}_{\circled{1}}\left\{ \underbrace{\frac{\ddot{W}^\prime\ddot{M}(\mathring{R}-\ddot{R})\delta}{N}}_{ \circled{2}}+\underbrace{\frac{\ddot{W}^\prime\ddot{M}(W^\ast-\ddot{W})\theta}{N}}_{\circled{3}}+\underbrace{\frac{\ddot{W}^\prime\ddot{M}\Delta\xi}{N}}_{\circled{4}}\right\} \nonumber \\
			&  =\underbrace{\left(\frac{\ddot{W}^\prime\ddot{M}\ddot{W}}{N}\right)^{-1}}_{\circled{1}}\left\{ \underbrace{\frac{\ddot{W}^\prime\ddot{M}\mathring{R}\delta}{N}}_{ \circled{2}}+\underbrace{\frac{\ddot{W}^\prime\ddot{M}(W^\ast-\ddot{W})\theta}{N}}_{\circled{3}}+\underbrace{\frac{\ddot{W}^\prime\ddot{M}\Delta\xi}{N}}_{\circled{4}}\right\}
	\end{align}}
	Let's first consider \circled{1} which is equal to
	{\small \begin{equation*}
			\begin{split}
				\frac{\ddot{W}^\prime (I-\ddot{R}(\ddot{R}^\prime \ddot{R})^{-1}\ddot{R}^\prime)\ddot{W}}{N}  &= \frac{\ddot{W}^\prime \ddot{W}}{N}-\frac{\ddot{W}^\prime\ddot{R}(\ddot{R}^\prime \ddot{R})^{-1}\ddot{R}^\prime\ddot{W}}{N} \\
				& = \frac{1}{N}\sum_{i=1}^{N}\ddot{W}_i^\prime \ddot{W}_i-\left(\frac{1}{N}\sum_{i=1}^{N}\ddot{W}_i^\prime\ddot{R}_i\right)\left(\frac{1}{N}\sum_{i=1}^{N}\ddot{R}_i^\prime\ddot{R}_i\right)^{-1}\left(\frac{1}{N}\sum_{i=1}^{N}\ddot{R}_i^\prime\ddot{W_i}\right)
			\end{split}
	\end{equation*}} Define $\dot{R}_i = (1, \dot{X}_i)$ where $\dot{X}_i = X_i-\mathbb{E}(X_i|D_i=1)$ and $\dot{W}_i = D_i\dot{R}_i$. Now since $\bar{X}_1\overset{p}{\rightarrow} \mathbb{E}(X|D=1)$, this implies that $\ddot{R}_i \overset{p}{\rightarrow} \dot{R}_i$ and $\ddot{W}_i \overset{p}{\rightarrow} \dot{W}_i$. Therefore, 
	{\small \begin{equation}\label{t1}
			\circled{1} \overset{p}{\rightarrow} \left\{\mathbb{E}(\dot{W}_i^\prime \dot{W}_i)-\mathbb{E}(\dot{W}_i^\prime\dot{R}_i)\mathbb{E}(\dot{R}_i^\prime\dot{R}_i)^{-1}\mathbb{E}(\dot{R}_i^\prime\dot{W}_i)\right\}^{-1}
	\end{equation}}
	Consider \circled{2} which is equal to 
	{\small \begin{align}\label{t2}
			\frac{\ddot{W}^\prime\ddot{M}\mathring{R}\delta}{N} &= \frac{\ddot{W}^\prime(I-\ddot{R}(\ddot{R}^\prime \ddot{R})^{-1}\ddot{R}^\prime)\mathring{R}\delta}{N} \nonumber \\
			& = \frac{\ddot{W}^\prime\mathring{R}\delta}{N} -\frac{\ddot{W}^\prime\ddot{R}(\ddot{R}^\prime \ddot{R})^{-1}\ddot{R}^\prime\mathring{R}\delta}{N}  \nonumber \\
			& = \left\{\frac{1}{N}\sum_{i=1}^{N}\ddot{W}_i^\prime\mathring{R}_i- \left(\frac{1}{N}\sum_{i=1}^{N}\ddot{W}_i^\prime\ddot{R}_i\right)\left(\frac{1}{N}\sum_{i=1}^{N}\ddot{R}_i^\prime\ddot{R}_i\right)^{-1}\left(\frac{1}{N}\sum_{i=1}^{N}\ddot{R}_i^\prime\mathring{R}_i\right)\right\}\delta \nonumber \\
			& \overset{p}{\rightarrow} \left\{\mathbb{E}(\dot{W}_i^\prime\mathring{R}_i)- \mathbb{E}(\dot{W}_i^\prime\dot{R}_i)[\mathbb{E}(\dot{R}_i^\prime\dot{R}_i)]^{-1}\mathbb{E}(\dot{R}_i^\prime\mathring{R}_i)\right\}\delta
	\end{align}} 
	Then, consider \circled{3} which is equal to 
	{\small \begin{align}\label{t3}
			&\frac{\ddot{W}^\prime(I-\ddot{R}(\ddot{R}^\prime\ddot{R})^{-1}\ddot{R}^\prime)(W^\ast-\ddot{W})\theta}{N} \nonumber \\
			& = \left\{\frac{\ddot{W}^\prime(W^\ast-\ddot{W})}{N}-\frac{\ddot{W}^\prime\ddot{R}(\ddot{R}^\prime\ddot{R})^{-1}\ddot{R}^\prime(W^\ast-\ddot{W})}{N} \right\} \theta \nonumber \\
			& \overset{p}{\rightarrow} \left\{\mathbb{E}[\dot{W}_i^\prime(W_i^\ast-\dot{W}_i)] - \mathbb{E}(\dot{W}_i^\prime\dot{R}_i)[\mathbb{E}(\dot{R}_i^\prime \dot{R}_i)]^{-1}\mathbb{E}[\dot{R}_i^\prime(W_i^\ast-\dot{W}_i)]\right\}\theta
	\end{align}}
	Finally, \circled{4} is equal to 
	{\small\begin{align}\label{t4}
			\frac{\ddot{W}^\prime(I-\ddot{R}(\ddot{R}^\prime\ddot{R})^{-1}\ddot{R}^\prime)\Delta\xi}{N} &= \frac{\ddot{W}^\prime\Delta\xi}{N}-\frac{\ddot{W}^\prime\ddot{R}(\ddot{R}^\prime\ddot{R})^{-1}\ddot{R}^\prime\Delta\xi}{N} \nonumber  \\
			& \overset{p}{\rightarrow} \mathbb{E}(\dot{W}_i^\prime\Delta\xi_i)-\mathbb{E}(\dot{W}_i^\prime\dot{R}_i)[\mathbb{E}(\dot{R}_i^\prime \dot{R}_i)]^{-1}\mathbb{E}(\dot{R}_i^\prime\Delta\xi_i) \nonumber  \\
			& = \mathbb{E}(\dot{W}_i^\prime\Delta\xi_i)
	\end{align}} since $\mathbb{E}(\dot{R}_i^\prime\Delta\xi_i) = \mathbb{E}[(1, \dot{X}_i)^\prime\Delta\xi_i] = [\mathbb{E}(\Delta\xi_i), \mathbb{E}(\dot{X}_i\Delta\xi_i)]^\prime = 0$. 
	
	Consider, 
	{\small \begin{align*}
			\mathbb{E}[\dot{W}_i^\prime \Delta\xi_i] &= \mathbb{E}[D_i\dot{R}_i^\prime \Delta\xi_i] \\
			& =  \mathbb{E}[\mathbbm{1}(R_i\gamma+U_i\geq 0, Z_i\alpha+V_i \geq 0)\dot{R}_i^\prime\Delta\xi_i] \\
			& = \mathbb{E}\left[\dot{R}_i^\prime \cdot \mathbb{E}\left( \Delta\xi_i|U_i\geq -R_i\gamma, V_i\geq -Z_i\alpha\right)\cdot\mathbb{P}(U_i\geq -R_i\gamma, V_i \geq -Z_i\alpha)\right] 
	\end{align*}}
	Then by using the general formulas for truncated normal distributions in \citet{tallis1961moment}, we can reduce the above expression by integration to 
	\begin{equation*}
		\begin{split}
			&\mathbb{E}\left(\Delta\xi_i|U_i\geq -R_i\gamma, V_i\geq -Z_i\alpha\right)\cdot \mathbb{P}(U_i\geq -R_i\gamma, V_i \geq -Z_i\alpha) \\
			& = \sigma\psi_v \phi(-Z_i\alpha)\Phi\left(\frac{R_i\gamma-\rho Z_i\alpha}{\sqrt{1-\rho^2}}\right) 
		\end{split}
	\end{equation*} Then, 
	\begin{equation}\label{tallis}
		\mathbb{E}[\dot{W}_i^\prime \Delta\xi_i] = \sigma\psi_v\mathbb{E}\left[\dot{R}_i^\prime  \phi(-Z_i\alpha)\Phi\left(\frac{R_i\gamma-\rho Z_i\alpha}{\sqrt{1-\rho^2}}\right)\right]
	\end{equation}
	
	Therefore, using \ref{t1}, \ref{t2}, \ref{t3}, \ref{t4}, and \ref{tallis} in the OLS bias expression in \ref{ols_bias}, we obtain the desired result, 
	\begin{equation}
		\text{plim}(\hat{\theta}_{FD})-\theta = Q^{-1}(A\delta+B\theta+C)
	\end{equation}
	where 
	\begin{align*}
		Q& =  \mathbb{E}(\dot{W}_i^\prime \dot{W}_i)-\mathbb{E}(\dot{W}_i^\prime\dot{R}_i)\mathbb{E}(\dot{R}_i^\prime\dot{R}_i)^{-1}\mathbb{E}(\dot{R}_i^\prime\dot{W}_i) \\
		A &=  \mathbb{E}(\dot{W}_i^\prime\mathring{R}_i)- \mathbb{E}(\dot{W}_i^\prime\dot{R}_i)[\mathbb{E}(\dot{R}_i^\prime\dot{R}_i)]^{-1}\mathbb{E}(\dot{R}_i^\prime\mathring{R}_i) \\
		B & = \mathbb{E}[\dot{W}_i^\prime(W_i^\ast-\dot{W}_i)] - \mathbb{E}(\dot{W}_i^\prime\dot{R}_i)[\mathbb{E}(\dot{R}_i^\prime \dot{R}_i)]^{-1}\mathbb{E}[\dot{R}_i^\prime(W_i^\ast-\dot{W}_i) \\ 
		C & = \sigma\psi_v\mathbb{E}\left[\dot{R}_i^\prime \phi(-Z_i\alpha)\Phi\left(\frac{R_i\gamma-\rho Z_i\alpha}{\sqrt{1-\rho^2}}\right)\right]
	\end{align*}
	Now since, $\hat{\tau}_{FD} = \frac{1}{N_1}\sum_{i:D_i=1}\ddot{R}_i\hat{\theta}_{FD} = \frac{1}{N_1}\sum_{i=1}^{N}D_i\ddot{R}_i\hat{\theta}_{FD}$, therefore, 
	\begin{equation*}
		\begin{split}
			\hat{\tau}_{FD}-\tau &= \frac{1}{N_1}\sum_{i=1}^{N}D_i\ddot{R}_i\hat{\theta}_{FD}-\tau = \frac{N}{N_1}\cdot \frac{1}{N}\sum_{i=1}^{N}D_i\ddot{R}_i\cdot \left(\hat{\theta}_{FD}-\theta+\theta\right)-\tau \\
			\text{ Note that } &\frac{N}{N_1} =[\mathbb{P}(D_i=1)]^{-1} + o_p(1) \text{ and  since }  \bar{X}_1 \overset{p}{\rightarrow} \mathbb{E}[X_i|D_i=1], \text{ this implies that }\\ 
			\hat{\tau}_{FD}-\tau & = \{\mathbb{P}(D_i=1)\}^{-1}\cdot \left\{ \frac{1}{N}\sum_{i=1}^{N}D_i\dot{R}_i\theta +  \frac{1}{N}\sum_{i=1}^{N}D_i\dot{R}_i\left(\hat{\theta}_{FD}-\theta\right) \right\} -\tau +o_p(1) 
		\end{split}
	\end{equation*} This implies, 
	\begin{equation*}
		\begin{split}
			\text{plim}(	\hat{\tau}_{FD})-\tau  &= \mathbb{E}[\dot{R}_i\theta|D_i=1]+ \mathbb{E}[\dot{R}_i|D_i=1]\cdot \text{plim}(\hat{\theta}_{FD}-\theta) -\tau  \\
			& = \tau+\mathbb{E}[\dot{R}_i  Q^{-1}(A\delta+B\theta+C)|D_i=1]-\tau \\
			& = \mathbb{E}[\dot{R}_i  Q^{-1}(A\delta+B\theta+C)|D_i=1]
		\end{split}	 
	\end{equation*}
\end{proof}

\begin{proof}(Bias of $\hat{\theta}_{POLS}$)
	Consider the following separate regression for the $T_i=1$ sample,
	\begin{equation*}
		Y_i \text{ on } \ddot{R}_i, \ddot{W}_i \ \text{ for } \ T_i =1 
	\end{equation*} where we are interested in estimating the coefficient on $W_i$. Let $Y_1$, $\ddot{R}_1$, $\ddot{W}_1$, and $\epsilon_1$ represent the $N$-vector data matrices with the $i$-th element given by $Y_{i1} = T_iY_i$, $\ddot{R}_{i1} = T_i\ddot{R}_i$, $\ddot{W}_{i1} = T_i\ddot{W}_i$, and $\epsilon_{i1} = T_i\epsilon_i$ respectively. We can then use Frisch-Waugh to obtain
	\begin{align*}
		\ddot{M}_{1} Y_1  =   \ddot{M}_{1}  \ddot{W}_1\pi_2+ \ddot{M}_{1} \epsilon_1  
	\end{align*} where $\pi_2 = \eta_2+\theta$, $\ddot{M}_1 = I_1-\ddot{R}_1(\ddot{R}_1^\prime \ddot{R}_1)^{-1}\ddot{R}_1^\prime$ is the residual making matrix for $\ddot{R}_1$. Also, $I_1$ is the matrix with $i$-th diagonal elements given by $T_i$. 
	\begin{align*}
		\hat{\pi}_2& = \left(\ddot{W}_1^\prime \ddot{M}_1 \ddot{W}_1\right)^{-1}\ddot{W}_1^\prime \ddot{M}_1Y_1 \\
		& = \pi_2+ \left(\ddot{W}_1^\prime \ddot{M}_1 \ddot{W}_1\right)^{-1}\ddot{W}_1^\prime \ddot{M}_1\epsilon_1
	\end{align*} which implies
	{\small \begin{align}\label{bias_rc}
			\hat{\pi}_2-\pi_2 &= \left(\frac{\ddot{W}_1^\prime \ddot{M}_1 \ddot{W}_1}{N}\right)^{-1}\frac{\ddot{W}_1^\prime\ddot{M}_1\epsilon_1}{N} \nonumber  \\
			& = \left(\frac{\ddot{W}_1^\prime \ddot{M}_1 \ddot{W}_1}{N}\right)^{-1}\frac{\ddot{W}_1^\prime\ddot{M}_1}{N}\{(\mathring{R}_1-\ddot{R}_1)\pi_1+(W_1^\ast-\ddot{W}_1)\pi_2+\xi^1\} \nonumber  \\
			& =  \left(\frac{\ddot{W}_1^\prime \ddot{M}_1 \ddot{W}_1}{N}\right)^{-1}\left\{\frac{\ddot{W}_1^\prime\ddot{M}_1\mathring{R}_1\pi_1}{N}+\frac{\ddot{W}_1^\prime\ddot{M}_1(W_1^\ast-\ddot{W}_1)\pi_2}{N}+\frac{\ddot{W}_1^\prime\ddot{M}_1\xi^1}{N}\right\} 
	\end{align}}
	 where $\mathring{R}_1$, $W_1^\ast$, and $\xi^1$ are again data matrices with $i$-th element given by $\mathring{R}_{i1} = T_i\mathring{R}_{i}$, $W_{i1}^\ast = T_iW_{i}^\ast$, and $\xi^1_{i} = T_i\xi_i = T_i\xi_{i1}$ respectively. Also, $\pi_1 = \eta_1+\delta$. 
	 
	Following the proof as in the case of the two period panel, we get
	{\small 
		\begin{align}\label{t0_rc}
			\frac{\ddot{W}_1^\prime \ddot{M}_1 \ddot{W}_1}{N}& = \frac{\ddot{W}_1^\prime(I_1-\ddot{R}_1(\ddot{R}_1^\prime \ddot{R}_1)^{-1}\ddot{R}_1^\prime)\ddot{W}_1}{N} \nonumber \\
			& =  \frac{\ddot{W}_1^\prime I_1\ddot{W}_1}{N}-\frac{\ddot{W}_1^\prime \ddot{R}_1(\ddot{R}_1^\prime \ddot{R}_1)^{-1}\ddot{R}_1^\prime\ddot{W}_1}{N} \nonumber\\
			& = \frac{1}{N}\sum_{i=1}^{N}T_i\ddot{W}_i^\prime \ddot{W}_i - \left(\frac{1}{N}\sum_{i=1}^{N}T_i\ddot{W}_i^\prime\ddot{R}_i\right)\left(\frac{1}{N}\sum_{i=1}^{N}T_i\ddot{R}_i^\prime\ddot{R}_i\right)^{-1}\left(\frac{1}{N}\sum_{i=1}^{N}T_i\ddot{R}_i^\prime\ddot{W}_i\right) \nonumber \\
			&\overset{p}{\longrightarrow} \lambda \cdot \left\{\mathbb{E}(\dot{W}_i^\prime \dot{W}_i|T_i=1)-\mathbb{E}(\dot{W}_i^\prime\dot{R}_i|T_i=1)[\mathbb{E}(\dot{R}_i^\prime \dot{R}_i)|T_i=1]^{-1}\mathbb{E}(\dot{R}_i^\prime \dot{W}_i|T_i=1)\right\}
	\end{align}}
	and, 
	{\small \begin{align}\label{t1_rc}
			\begin{split}
				\frac{\ddot{W}_1^\prime\ddot{M}_1\mathring{R}_1\pi_1}{N} &\overset{p}{\longrightarrow} \lambda \left\{\mathbb{E}(\dot{W}_i^\prime\mathring{R}_i|T_i=1)-\mathbb{E}(\dot{W}_i^\prime\dot{R}_i|T_i=1)[\mathbb{E}(\dot{R}_i^\prime\dot{R}_i|T_i=1)]^{-1}\mathbb{E}(\dot{R}_i^\prime\mathring{R}_i|T_i=1)\right\}\pi_1 \\
				\frac{\ddot{W}_1^\prime\ddot{M}_1(W_1^\ast-\ddot{W}_1)\pi_2}{N} &\overset{p}{\longrightarrow} \lambda\bigg\{\mathbb{E}[\dot{W}_i^\prime(W_i^\ast-\dot{W}_i)|T_i=1] - \mathbb{E}(\dot{W}_i^\prime\dot{R}_i|T_i=1)[\mathbb{E}(\dot{R}_i^\prime\dot{R}_i|T_i=1)]^{-1}\\
				&\mathbb{E}[\dot{R}_i^\prime(W_i^\ast-\dot{W}_i|T_i=1)] \bigg\} \pi_2 \\
				\frac{\ddot{W}_1^\prime\ddot{M}_1\xi^1}{N}& \overset{p}{\longrightarrow} \lambda \  \left\{\mathbb{E}(\dot{W}_i^\prime\xi_{i1}|T_i=1)-\mathbb{E}(\dot{W}_i^\prime\dot{R}_i|T_i=1)[\mathbb{E}(\dot{R}_i^\prime\dot{R}_i|T_i=1)]^{-1}\mathbb{E}(\dot{R}_i^\prime\xi_{i1}|T_i=1)\right\} \\
				&= \lambda\cdot \mathbb{E}(\dot{W}_i^\prime\xi_{i1}|T_i=1) 
			\end{split}
	\end{align}}
	Similarly, let $Y_0$, $\ddot{R}_0$, $\ddot{W}_0$, $\epsilon_0$ be again data matrices with the $i$-th element given by $Y_{i0} = (1-T_i)Y_i$, $\ddot{R}_{i0} = (1-T_i)\ddot{R}_i$, $\ddot{W}_{i0} = (1-T_i)\ddot{W}_i$, and $\epsilon_{i0} = (1-T_i)\epsilon_{i}$, respectively. Then, using Frisch-Waugh, 
	\begin{align}
		\ddot{M}_0Y_0 = \ddot{M}_0\ddot{W}_0\eta_2+\ddot{M}_0\epsilon_0
	\end{align} where $\ddot{M}_0 = (I_0-\ddot{R}_0(\ddot{R}^\prime_0\ddot{R}_0)^{-1}\ddot{R}^\prime_0)$ is the residual making matrix for $\ddot{R}_0$.
	Then the bias for $\hat{\eta}_2$ is given as
	{\small \begin{align}\label{bias2_rc}
			\hat{\eta}_2-\eta_2 
			& =  \left(\frac{\ddot{W}_0^\prime \ddot{M}_0 \ddot{W}_0}{N}\right)^{-1}\left\{\frac{\ddot{W}_0^\prime\ddot{M}_0\mathring{R}_0\eta_1}{N}+\frac{\ddot{W}_0^\prime\ddot{M}_0(W_0^\ast-\ddot{W}_0)\eta_2}{N}+\frac{\ddot{W}_0^\prime\ddot{M}_0\xi^0}{N}\right\} 
	\end{align}}
	where $\mathring{R}_0$, $W^\ast_0$, and $\xi^0$ are defined analogously and for each term we have
	{\small \begin{align}\label{t2_rc}
			\frac{\ddot{W}_0^\prime \ddot{M}_0 \ddot{W}_0}{N} &\overset{p}{\longrightarrow} (1-\lambda)\bigg\{\mathbb{E}(\dot{W}_i^\prime \dot{W}_i|T_i=0)-\mathbb{E}(\dot{W}_i^\prime\dot{R}_i|T_i=0)[\mathbb{E}(\dot{R}_i^\prime \dot{R}_i|T_i=0)]^{-1} \mathbb{E}(\dot{R}_i^\prime \dot{W}_i|T_i=0)\bigg\} \nonumber \\
			\frac{\ddot{W}_0^\prime\ddot{M}_0\mathring{R}_0\eta_1}{N} &\overset{p}{\longrightarrow} (1-\lambda)\left\{\mathbb{E}(\dot{W}_i^\prime\mathring{R}_i|T_i=0)-\mathbb{E}(\dot{W}_i^\prime\dot{R}_i|T_i=0)[\mathbb{E}(\dot{R}_i^\prime\dot{R}_i|T_i=0)]^{-1}\mathbb{E}(\dot{R}_i^\prime\mathring{R}_i|T_i=0)\right\}\eta_1 \nonumber \\
			\frac{\ddot{W}_0^\prime\ddot{M}_0(W_0^\ast-\ddot{W}_0)\eta_2}{N} &\overset{p}{\longrightarrow} (1-\lambda)\bigg\{\mathbb{E}[\dot{W}_i^\prime(W_i^\ast-\dot{W}_i|T_i=0)] - \mathbb{E}(\dot{W}_i^\prime\dot{R}_i|T_i=0)[\mathbb{E}(\dot{R}_i^\prime\dot{R}_i|T_i=0)]^{-1} \nonumber \\
			&\mathbb{E}[\dot{R}_i^\prime(W_i^\ast-\dot{W}_i|T_i=0)] \bigg\}\eta_2 \nonumber \\
			\frac{\ddot{W}_0^\prime\ddot{M}_0\xi^0}{N}& \overset{p}{\longrightarrow} (1-\lambda) \  \mathbb{E}(\dot{W}_i^\prime\xi_{i0}|T_i=0)
	\end{align}}
	Then together with equations \ref{bias_rc}, \ref{t1_rc}, \ref{bias2_rc}, and \ref{t2_rc} we obtain
	\begin{align*}
		\text{Bias}(\hat{\theta}_{POLS}) &= \text{Bias}(\hat{\pi}_2)-\text{Bias}(\hat{\eta}_2) \\
		& = Q_1^{-1}\left(A_1\pi_1+B_1\pi_2+C_1\right)-Q_0^{-1}\left(A_0\eta_1+B_0\eta_2+C_0\right)
	\end{align*}
	where all the above objects are defined analogously to the case for $\hat{\theta}_{FD}$. The asymptotic bias expression for $\hat{\tau}_{POLS}$ can be derived analogously to the case for $\hat{\tau}_{FD}$.
\end{proof}

\begin{proof}(Consistency of $\hat{\theta}_{FD}^{2S}$) 
	{\small \begin{align}\label{consistency_tp}
			\hat{\theta}_{FD}^{2S}- \theta& = \left(\frac{\hat{W}^{\ast\prime} \hat{M}^\ast\hat{W}^\ast}{N}\right)^{-1}\frac{\hat{W}^{\ast\prime}\hat{M}^\ast\Delta\varepsilon}{N} \nonumber \\
			& = \left(\frac{\hat{W}^{\ast\prime} \hat{M}^\ast\hat{W}^\ast}{N}\right)^{-1}\frac{\hat{W}^{\ast\prime}\hat{M}^\ast}{N} \left[(\mathring{R}-\hat{R}^\ast)\delta+(W^\ast-\hat{W}^\ast)\theta+\Delta\xi\right] \nonumber \\
			& =  \left(\frac{\hat{W}^{\ast\prime} \hat{M}^\ast\hat{W}^\ast}{N}\right)^{-1} \left[\frac{\hat{W}^{\ast\prime}\hat{M}^\ast\mathring{R}\delta}{N}+\frac{\hat{W}^{\ast\prime}\hat{M}^\ast (W^\ast-\hat{W}^\ast)\theta}{N}+\frac{\hat{W}^{\ast\prime}\hat{M}^\ast\Delta\xi}{N}\right] 
	\end{align}}
	Now $\hat{R}^\ast \overset{p}{\longrightarrow} \mathring{R}$ because 
	\begin{equation*}
		\begin{split}
				\hat{\bar{X}}^\ast_1 = \frac{1}{N}\cdot\frac{N}{\hat{N}^\ast} \sum_{i=1}^{N}\hat{D}_i^\ast\cdot X_i &= \frac{1}{\mathbb{P}(D_i^\ast =1)}\cdot \frac{1}{N}\sum_{i=1}^{N}\hat{D}_i^\ast\cdot X_i +o_p(1) \\
				& = \frac{1}{\mathbb{P}(D_i^\ast =1)}\cdot \mathbb{E}(D_i^\ast X_i)+o_p(1) \\
				& = \mathbb{E}(X_i|D_i^\ast=1)+o_p(1)
		\end{split}	
	\end{equation*} where the second equality follows from the fact that $\hat{D}^\ast = \Phi(R\hat{\gamma})\overset{p}{\longrightarrow} \Phi(R\gamma)$. This implies that as $N\rightarrow \infty$, \[\hat{\bar{X}}_1^\ast \overset{p}{\longrightarrow}  \mathbb{E}(X_i|D_i^\ast=1) \]  Hence, 
	{\small \begin{align*}
			\left(\frac{\hat{W}^{\ast\prime} \hat{M}^\ast\hat{W}^\ast}{N}\right)^{-1} \overset{p}{\longrightarrow} \mathbb{E}[\Phi^2(R_i\gamma)\mathring{R}_i^\prime\mathring{R}_i]-\mathbb{E}[\Phi(R_i\gamma)\mathring{R}_i^\prime\mathring{R}_i][\mathbb{E}(\mathring{R}_i^\prime\mathring{R}_i)]^{-1}\mathbb{E}[\Phi(R_i\gamma)\mathring{R}_i^\prime\mathring{R}_i]
	\end{align*} }
	and
	{\small \begin{align}
			\frac{\hat{W}^{\ast\prime}\hat{M}^\ast\mathring{R}\delta}{N} &\overset{p}{\longrightarrow} \left\{\mathbb{E}[\Phi(R_i\gamma)\mathring{R}_i^\prime\mathring{R}_i]-\mathbb{E}[\Phi(R_i\gamma)\mathring{R}_i^\prime\mathring{R}_i][\mathbb{E}(\mathring{R}_i^\prime\mathring{R}_i)]^{-1}\mathbb{E}[\mathring{R}_i^\prime\mathring{R}_i]\right\}\delta  \nonumber \\
			&= 0 
	\end{align}}
	Now, 
	{\small \begin{align}
			\frac{\hat{W}^{\ast\prime}\hat{M}^\ast (W^\ast-\hat{W}^\ast)\theta}{N}&\overset{p}{\longrightarrow} \bigg\{\mathbb{E}[\Phi(R_i\gamma)\mathring{R}_i^\prime(D_i^\ast-\Phi(R_i\gamma))\mathring{R}_i]-\mathbb{E}[\Phi(R_i\gamma)\mathring{R}_i^\prime\mathring{R}_i][\mathbb{E}(\mathring{R}_i^\prime\mathring{R}_i)]^{-1} \nonumber \\
			&\mathbb{E}[\mathring{R}_i^\prime(D_i^\ast-\Phi(R_i\gamma))\mathring{R}_i]\bigg\}\theta \nonumber \\
			& = 0
	\end{align}}
	where the last equality is due to  law of iterated expectations because $\mathbb{E}[D_i^\ast-\Phi(R_i\gamma)|R_i] = 0$. Finally, 
	{\small\begin{equation*}
			\begin{split}
				\frac{\hat{W}^{\ast\prime}\hat{M}^\ast\Delta\xi}{N} &\overset{p}{\longrightarrow} \mathbb{E}[\Phi(R_i\gamma)\mathring{R}_i^\prime \Delta\xi_i]-\mathbb{E}[\Phi(R_i\gamma)\mathring{R}_i^\prime\mathring{R}_i][\mathbb{E}(\mathring{R}_i^\prime \mathring{R}_i)]^{-1}\mathbb{E}[\mathring{R}_i^\prime \Delta\xi_i] \\
				& = 0
			\end{split}
	\end{equation*}} where again the last equality follows due to $\mathbb{E}(\mathring{R}_i^\prime\Delta\xi_i) = 0$ and law of iterated expectations. 
	
	Therefore, $\hat{\theta}_{FD}^{2S}-\theta \overset{p}{\longrightarrow} 0$. 	Now, 
	{\small\begin{equation*}
			\begin{split}
				\hat{\tau}^{2S}_{FD} &= \frac{1}{\hat{N}^\ast}\sum_{i=1}^{N}\hat{D}^\ast_i\hat{R}^\ast_i\hat{\theta}^{2S}_{FD}  = \frac{N}{\hat{N}^\ast}\left\{\frac{1}{N} \sum_{i=1}^{N}\hat{D}^\ast_i\hat{R}^\ast_i\theta+ \frac{1}{N} \sum_{i=1}^{N}\hat{D}^\ast_i\hat{R}^\ast_i\left(\hat{\theta}^{2S}_{FD} -\theta \right) \right\}
			\end{split}
	\end{equation*}}
	Now since  $\frac{\hat{N}^\ast}{N} = \mathbb{P}(D_i^\ast=1)+o_p(1)$ and $\frac{1}{N}\sum_{i=1}^{N}\hat{D}_i^\ast \hat{R}_i^\ast=  \frac{1}{N}\sum_{i=1}^{N}\Phi(R_i\gamma)\mathring{R}_i+o_p(1)=  \mathbb{E}[\Phi(R_i\gamma)\mathring{R}_i] + o_p(1)$. %
	Therefore,
	\begin{equation*}
		\hat{\tau}^{2S}_{FD} = \frac{1}{\mathbb{P}(D_i^\ast=1)}\left\{\frac{1}{N}\sum_{i=1}^{N}\Phi(R_i\gamma)\mathring{R}_i\theta+ \frac{1}{N} \sum_{i=1}^{N}\Phi(R_i\gamma)\mathring{R}_i\left(\hat{\theta}^{2S}_{FD} -\theta \right) \right\}+o_p(1) 
	\end{equation*}
	Now, $\frac{1}{N}\sum_{i=1}^{N}\Phi(R_i\gamma)\mathring{R}_i\theta=\mathbb{E}[\mathbb{P}(D_i^\ast=1|R_i)\mathring{R}_i]+o_p(1)$ where $\mathbb{E}[\mathbb{P}(D_i^\ast=1|R_i)\mathring{R}_i]= \mathbb{E}[\mathbb{E}(D_i^\ast\mathring{R}_i|R_i)] = \mathbb{E}[D_i^\ast\mathring{R}_i]$. Hence, 
	\begin{equation*}
		\text{plim}(\hat{\tau}^{2S}_{FD}-\tau) =\mathbb{E}[\mathring{R}_i\cdot\text{plim}(\hat{\theta}^{2S}_{FD} -\theta)|D_i^\ast=1] = 0
	\end{equation*}
	Therefore, $\hat{\tau}^{2S}_{FD}-\tau \overset{p}{\rightarrow} 0$.
\end{proof}

\begin{proof}(Consistency of $\hat{\theta}^{2S}_{POLS}$) Let $\hat{\pi}^{2S}_2$ be the two-step estimator of the coefficient on $\hat{W}^\ast_i$ from estimating the regression of \[Y_i \text{ on } \hat{R}^\ast_i, \hat{W}^\ast_i \text{ if } T_i=1 \ \text{ for } i=1,\ldots, N \] which again using Frisch-Waugh gives us the following estimating equation
	\begin{align}
		 \hat{M}^\ast_1 Y_1 =  \hat{M}^\ast_1\hat{W}_1^\ast + \hat{M}^\ast_1\varepsilon_1
	\end{align} where $\hat{M}^\ast_1 = I_1-\hat{R}_1^\ast(\hat{R}_1^{\ast\prime} \hat{R}_1^\ast)^{-1}\hat{R}_1^{\ast\prime}$ is the residual making matrix for $\hat{R}^\ast_1$ whose $i$-th element is given as $T_i\hat{R}_i^\ast$ and $\hat{W}_1^\ast$, $\varepsilon_1$ are defined analogously.
	
	Then, 
	{\small \begin{align}\label{bias_pi2} 
			\hat{\pi}_2^{2S}- \pi_2& = \left(\frac{\hat{W}_1^{\ast\prime} \hat{M}^\ast_1\hat{W}_1^\ast}{N}\right)^{-1}\frac{\hat{W}_1^{\ast\prime}\hat{M}^\ast_1\varepsilon_1}{N} \nonumber \\
			& =  \left(\frac{\hat{W}_1^{\ast\prime} \hat{M}_1^\ast\hat{W}_1^\ast}{N}\right)^{-1}\frac{\hat{W}_1^{\ast\prime}\hat{M}^\ast_1}{N}\left\{(\mathring{R}_1-\hat{R}_1^\ast)\pi_1+(W_1^\ast-\hat{W}_1^\ast)\pi_2+\xi^1\right\} \nonumber \\
			& = \left(\frac{\hat{W}_1^{\ast\prime} \hat{M}^\ast_1\hat{W}_1^\ast}{N}\right)^{-1}\left\{\frac{\hat{W}_1^{\ast\prime}\hat{M}^\ast_1\mathring{R}_1\pi_1}{N}+\frac{\hat{W}_1^{\ast\prime}\hat{M}^\ast_1(W_1^\ast-\hat{W}_1^\ast)\pi_2}{N}+\frac{\hat{W}_1^{\ast\prime}\hat{M}^\ast_1\xi^1}{N}\right\} 
	\end{align}}
	where 
	{\small \begin{align}\label{t1pi2}
			\left(\frac{\hat{W}_1^{\ast\prime} \hat{M}^\ast_1\hat{W}_1^\ast}{N}\right)^{-1} \overset{p}{\longrightarrow} \lambda \left\{ \mathbb{E}[\Phi^2(R_i\gamma)\mathring{R}_i^\prime \mathring{R}_i] - \mathbb{E}[\Phi(R_i\gamma)\mathring{R}_i^\prime\mathring{R}_i][\mathbb{E}(\mathring{R}_i^\prime \mathring{R}_i)]^{-1}\mathbb{E}[\Phi(R_i\gamma)\mathring{R}_i^\prime \mathring{R}_i]\right\}	
	\end{align}}
	and 
	{\small \begin{align}\label{t2pi2}
			&\frac{\hat{W}_1^{\ast\prime}\hat{M}^\ast_1\mathring{R}_1\pi_1}{N} = \frac{\hat{W}_1^{\ast\prime}\{I_1-\hat{R}_1^\ast(\hat{R}_1^{\ast\prime}\hat{R}_1^\ast)^{-1}\hat{R}_1^{\ast\prime}\}\mathring{R}_1\pi_1}{N} \nonumber \\
			& = \left\{\frac{\hat{W}_1^{\ast\prime}I_1\mathring{R}_1}{N}-\frac{\hat{W}_1^{\ast\prime}\hat{R}_1^\ast(\hat{R}_1^{\ast\prime}\hat{R}_1^\ast)^{-1}\hat{R}_1^{\ast\prime}\mathring{R}_1}{N}\right\}\pi_1 \nonumber \\
			& = \left\{\frac{1}{N}\sum_{i=1}^{N}T_i\hat{W}_i^{\ast\prime}\mathring{R}_i-\left(\frac{1}{N}\sum_{i=1}^{N}T_i\hat{W}_i^{\ast\prime}\hat{R}^\ast_i\right) \left(\frac{1}{N}\sum_{i=1}^{N}T_i\hat{R}_i^{\ast\prime}\hat{R}_i^\ast\right)^{-1}\left(\frac{1}{N}\sum_{i=1}^{N}T_i\hat{R}_i^{\ast\prime}\mathring{R}_i\right)\right\}\pi_1 \nonumber \\
			&\overset{p}{\longrightarrow} \lambda\left\{\mathbb{E}[\Phi(R_i\gamma)\mathring{R}_i^\prime\mathring{R}_i] - \mathbb{E}[\Phi(R_i\gamma)\mathring{R}_i^\prime\mathring{R}_i][\mathbb{E}(\mathring{R}_i^\prime\mathring{R}_i)]^{-1}\mathbb{E}[\mathring{R}_i^\prime\mathring{R}_i]\right\}\pi_1 \nonumber \\
			& = 0
	\end{align}}
	Similarly, 
	{\small \begin{align} \label{t3pi2}
			\frac{\hat{W}_1^{\ast\prime}\hat{M}^\ast_1(W_1^\ast-\hat{W}_1^\ast)\pi_2}{N} &\overset{p}{\longrightarrow} \lambda \bigg\{ \mathbb{E}[\Phi(R_i\gamma)\mathring{R}_i^\prime (D_i^\ast-\Phi(R_i\gamma))\mathring{R}_i]-\mathbb{E}[\Phi(R_i\gamma)\mathring{R}_i^\prime \mathring{R}_i][\mathbb{E}(\mathring{R}_i^\prime \mathring{R}_i)]^{-1}\cdot \nonumber \\
			& \mathbb{E}[\mathring{R}_i^\prime (D_i^\ast-\Phi(R_i\gamma))\mathring{R}_i]\bigg\}\pi_2 \nonumber \\
			& = 0 
	\end{align}}
	where the equality follows from the law of iterated expectations and the fact that $ D_i^\ast \overset{p}{\longrightarrow}  \Phi(R_i\gamma)$. 
	Finally, 
	{\small \begin{align}\label{t4pi2}
			\frac{\hat{W}_1^{\ast\prime}\hat{M}_1^\ast\xi^1}{N} & \overset{p}{\longrightarrow} \mathbb{E}[\Phi(R_i\gamma)\mathring{R}_i^\prime \xi_i]-\mathbb{E}[\Phi(R_i\gamma)\mathring{R}_i^\prime\mathring{R}_i][\mathbb{E}(\mathring{R}_i^\prime \mathring{R}_i)]^{-1}\mathbb{E}[\mathring{R}_i^\prime \xi_i] \nonumber \\
			& = 0
	\end{align}} where again the equality follows due to $\mathbb{E}(\mathring{R}_i^\prime\xi_i) = 0$ and law of iterated expectations. Therefore, together with \ref{bias_pi2}, \ref{t1pi2}, \ref{t2pi2}, \ref{t3pi2}, and \ref{t4pi2}, we obtain $\hat{\pi}^{2S}_2-\pi_2  \overset{p}{\longrightarrow} 0$. Following in a similar manner, let $\hat{\eta}^{2S}_2$ be the two-step estimator of the coefficient on $\hat{W}^\ast_i$ from estimating the regression of \[Y_i \text{ on } \hat{R}^\ast_i, \hat{W}^\ast_i \text{ if } T_i=0 \ \text{ for } i=1,\ldots, N \] Then, we can also show that $\hat{\eta}^{2S}_2-\eta_2  \overset{p}{\longrightarrow} 0$. Now since $\hat{\theta}_{POLS}^{2S} = \hat{\pi}^{2S}_2 - \hat{\eta}^{2S}_2$, therefore $\hat{\theta}^{2S}_{POLS}-\theta \overset{p}{\longrightarrow}0$. Similar to the case for the FD-estimator, we can show $\hat{\tau}^{2S}_{POLS}-\tau \overset{p}{\rightarrow}0$.
\end{proof}

 \begin{proof}(Asymptotic normality for $\hat{\theta}^{2S}_{FD}$)
	{\small \begin{equation*}
			\sqrt{N}(\hat{\theta}^{2S}_{FD}-\theta)  =  \left(\frac{\hat{W}^{\ast\prime} \hat{M}^\ast\hat{W}^\ast}{N}\right)^{-1} \frac{\hat{W}^{\ast\prime}\hat{M}^\ast}{\sqrt{N}} \left[(\mathring{R}-\hat{R}^\ast)\delta+(W^\ast-\hat{W}^\ast)\theta+\Delta\xi\right] 
	\end{equation*}}
	Now we have already shown that
	{\small	\begin{equation*}
			\begin{split}
				\frac{\hat{W}^{\ast\prime}\hat{M}^\ast\hat{W}^\ast}{N} &= \frac{1}{N}\sum_{i=1}^{N}\Phi^2(R_i\hat{\gamma})\hat{R}^{\ast\prime}_i\hat{R}^\ast_i-\left(\frac{1}{N}\sum_{i=1}^{N}\Phi(R_i\hat{\gamma})\hat{R}^{\ast\prime}_i\hat{R}^\ast_i\right)\left(\frac{1}{N}\sum_{i=1}^{N}\hat{R}^{\ast\prime}_i\hat{R}^\ast_i\right)^{-1}\left(\frac{1}{N}\sum_{i=1}^{N}\Phi(R_i\hat{\gamma})\hat{R}^{\ast\prime}_i\hat{R}^\ast_i\right) \\
				& = \frac{1}{N}\sum_{i=1}^{N}\hat{\Gamma}_i\hat{\Gamma}_i^\prime \overset{p}{\rightarrow} \mathbb{E}[\Gamma_i\Gamma_i^\prime] \equiv \Omega_{\Gamma}
			\end{split}
	\end{equation*}}
	where $\displaystyle \hat{\Gamma}_i = \Phi(R_i\hat{\gamma})\hat{R}^{\ast\prime}_i-\left(\frac{\sum_{i=1}^{N}\Phi(R_i\hat{\gamma})\hat{R}^{\ast\prime}_i\hat{R}^\ast_i}{N}\right)\left(\frac{\sum_{i=1}^{N}\hat{R}^{\ast\prime}_i\hat{R}^\ast_i}{N}\right)^{-1}\hat{R}^{\ast\prime}_i$ and $\displaystyle \Gamma_i = \Phi(R_i\gamma)\mathring{R}_i^\prime-\mathbb{E}[\Phi(R_i\gamma)\mathring{R}_i^\prime\mathring{R}_i]\mathbb{E}[\mathring{R}_i^\prime\mathring{R}_i]^{-1}\mathring{R}_i^\prime$. Therefore, using the above we can write
	{\small \begin{equation}\label{bias_theta_rc}
			\sqrt{N}(\hat{\theta}^{2S}_{FD}-\theta)  =  \Omega^{-1}_{\Gamma} \left(\frac{\hat{W}^{\ast\prime}\hat{M}^\ast}{\sqrt{N}} \left[(\mathring{R}-\hat{R}^\ast)\delta+(W^\ast-\hat{W}^\ast)\theta+\Delta\xi\right]\right) +o_p(1)
	\end{equation}}
	Define, 
	$\Pi_i^\ast(\gamma) = \Phi(R_i\gamma)\mathring{R}_i$ and let $\Pi^\ast(\gamma) = \left[\Pi_1^\ast(\gamma), \ldots, \Pi_N^\ast(\gamma)\right]^\prime$. Then, we can express the terms inside the bracket as 
	{\small \begin{align}\label{bias_theta_rc2}
				&\frac{\hat{W}^{\ast\prime}\hat{M}^\ast}{\sqrt{N}} \left[(\mathring{R}-\hat{R}^\ast)\delta+(W^\ast-\hat{W}^\ast -\Pi^\ast(\gamma)+\Pi^\ast(\gamma))\theta+\Delta\xi\right]  \nonumber \\
				& = \frac{\hat{W}^{\ast\prime}\hat{M}^\ast}{\sqrt{N}} \left[(\mathring{R}-\hat{R}^\ast)\delta+(W^\ast-\Pi^\ast(\gamma))\theta+
				(\Pi^\ast(\gamma)-\hat{W}^\ast)\theta+\Delta\xi\right] \nonumber \\
				& =  \frac{\hat{W}^{\ast\prime}\hat{M}^\ast\mathring{R}\delta}{\sqrt{N}}+\frac{\hat{W}^{\ast\prime}\hat{M}^\ast\{(W^\ast-\Pi^\ast(\gamma))\theta+\Delta\xi\}}{\sqrt{N}}+
				\frac{\hat{W}^{\ast\prime}\hat{M}^\ast(\Pi^\ast(\gamma)-\hat{W}^\ast)\theta}{\sqrt{N}} \nonumber  \\
				& = \sqrt{N}H_{1N}+\sqrt{N}H_{2N}+\sqrt{N}H_{3N}
	\end{align}}
	Then by the central limit theorem, 
	{\small \begin{align*}
				\sqrt{N}H_{1N} &= \frac{\hat{W}^{\ast\prime}\hat{M}^\ast \mathring{R}\delta}{\sqrt{N}} =  \frac{1}{\sqrt{N}}\sum_{i=1}^{N}\hat{\Gamma}_i\mathring{R}_i\delta = \frac{1}{\sqrt{N}}\sum_{i=1}^{N}\Gamma_i\mathring{R}_i\delta+o_p(1) \overset{d}{\longrightarrow}N(0,\Omega_{1\theta}) \\
				\Omega_{1\theta} &= \mathbb{E}[\Gamma_i\mathring{R}_i\delta\delta^\prime\mathring{R}_i^\prime\Gamma_i^\prime] \tag{since  $\mathbb{E}[\Gamma_i\mathring{R}_i] = 0$}
	\end{align*}}
	{\small \begin{align*}
		\sqrt{N}H_{2N} &= \frac{\hat{W}^{\ast\prime}\hat{M}^\ast[(W^\ast-\Pi^\ast(\gamma))\theta+\Delta\xi]}{\sqrt{N}} =  \frac{1}{\sqrt{N}}\sum_{i=1}^{N}\hat{\Gamma}_i\{(W^\ast_i-\Pi^\ast_i(\gamma))\theta+\Delta\xi_i\}\overset{d}{\longrightarrow}N(0,\Omega_{2\theta}) \\
		\Omega_{2\theta}&= \sigma^2\mathbb{E}[\Gamma_i\Gamma_i^\prime]+\mathbb{E}[\Gamma_i(W_i^\ast-\Pi_i^\ast(\gamma))\theta\theta^\prime(W_i^\ast-\Pi_i^\ast(\gamma))^\prime\Gamma_i^\prime]
	\end{align*}} due to  law of iterated expectations and the fact that $\mathbb{E}[\Delta\xi|D^\ast, X] = 0$ and
	{\small \begin{align*}
		\sqrt{N}H_{3N} =  \frac{\hat{W}^{\ast\prime}\hat{M}^\ast  (\Pi^\ast(\gamma)-\hat{W}^\ast)\theta}{\sqrt{N}} = \frac{1}{\sqrt{N}}\sum_{i=1}^{N}\hat{\Gamma}_i\{\Pi_i^\ast(\gamma)-\hat{W}_i^\ast\}\theta
\end{align*}}
	{\small \begin{align*}
		&=  \frac{1}{\sqrt{N}}\sum_{i=1}^{N}\Gamma_i\{\Pi_i^\ast(\gamma)-\hat{W}_i^\ast\}\theta+o_p(1) \\
		& = \frac{1}{\sqrt{N}}\sum_{i=1}^{N}\Gamma_i\{\Phi(R_i\gamma)\mathring{R}_i-\hat{D}_i^\ast(\hat{R}_i^\ast-\mathring{R}_i+\mathring{R}_i)\}\theta+o_p(1) \\
		& = \frac{1}{N}\sum_{i=1}^{N}\Gamma_i \cdot \{\sqrt{N}(\Phi(R_i\gamma)-\hat{D}_i^\ast)\mathring{R}_i-\Phi(R_i\gamma)\cdot \sqrt{N}(\hat{R}_i^\ast-\mathring{R}_i)\}\theta+o_p(1) 
	\end{align*}} 
	By the delta method, $\sqrt{N}(\Phi(R_i\gamma)-\hat{D}_i^\ast) \approx -\phi(R_i\tilde{\gamma})R_i\cdot \sqrt{N}(\hat{\gamma}-\gamma)$ where $\tilde{\gamma}$ lies between $\hat{\gamma}$ and $\gamma$. Since
	$(\hat{R}_i^\ast-\mathring{R}_i)\theta = -(\hat{\bar{X}}_1^\ast-\mu_1)\kappa$. 
	Hence, one may rewrite the above as
	{\small \begin{align*}
		\sqrt{N}H_{3N} &\approx \frac{1}{N}\sum_{i=1}^{N}-\Gamma_i \phi(R_i\tilde{\gamma})R_i\cdot \sqrt{N}(\hat{\gamma}-\gamma)\mathring{R}_i\theta -\frac{1}{N}\sum_{i=1}^{N}\Gamma_i\Phi(R_i\gamma)\cdot  \sqrt{N}(\hat{R}_i^\ast-\mathring{R}_i)\theta \\
		&=\frac{1}{N}\sum_{i=1}^{N}-\Gamma_i \phi(R_i\tilde{\gamma})R_i\mathring{R}_i\theta \cdot \sqrt{N}(\hat{\gamma}-\gamma)+\frac{1}{N}\sum_{i=1}^{N}\Gamma_i\Phi(R_i\gamma)\cdot  \sqrt{N}(\hat{\bar{X}}_1^\ast-\mu_1)\kappa 
	\end{align*}} where  $\sqrt{N}(\hat{\bar{X}}_1^\ast-\mu_1) \approx \frac{1}{\sqrt{N}}\sum_{i=1}^{N}\left(\frac{\Phi(R_i\gamma)X_i}{\rho}-\mu_1\right)+\frac{1}{N}\sum_{i=1}^{N}\frac{\Phi(R_i\gamma)R_i}{\rho}\cdot \sqrt{N}(\hat{\gamma}-\gamma)X_i$.
	 Now the covariance between $\sqrt{N}H_{1N}$ and $\sqrt{N}H_{2N}$ denoted by $
	\Omega_{12\theta} = 0$. Therefore, 
	\begin{equation*}
		\begin{split}
			\sqrt{N}(\hat{\theta}^{2S}_{FD}-\theta) &\overset{d}{\rightarrow} N(0, \Omega_{\Gamma}^{-1}\Omega_{\theta}\Omega_{\Gamma}^{-1}), \text{ where} \\
			\Omega_{\theta}& = \Omega_{1\theta}+\Omega_{2\theta} +\Omega_{3\theta}
		\end{split}
	\end{equation*}
Now, since 
{\small \begin{equation*}
	\sqrt{N}(\hat{\tau}_{FD}^{2S}-\tau) = \frac{1}{\mathbb{P}(D_i^\ast=1)}\cdot \frac{1}{N}\sum_{i=1}^{N}\Phi(R_i\gamma)\mathring{R}_i\cdot \sqrt{N}(\hat{\theta}_{FD}^{2S}-\theta)+o_p(1)
\end{equation*}}
Therefore, 
{\small\begin{equation*}
	\text{Avar}[\sqrt{N}(\hat{\tau}_{FD}^{2S}-\tau)] = \mathbb{E}(\mathring{R}_i|D_i^\ast=1)\cdot \text{Avar}[\sqrt{N}(\hat{\theta}_{FD}^{2S}-\theta)]\cdot \mathbb{E}(\mathring{R}_i|D_i^\ast=1)^\prime
		\end{equation*}}
\end{proof}

\begin{proof}(Asymptotic Normality of $\hat{\theta}_{2S}^{POLS}$) We know that 
	{\small \begin{align*}
			\sqrt{N}(\hat{\pi}_{2S}^{POLS}- \pi_2)
			& =  \left(\frac{\hat{W}_1^{\ast\prime} \hat{M}^\ast_1\hat{W}_1^\ast}{N}\right)^{-1}\frac{\hat{W}_1^{\ast\prime}\hat{M}^\ast_1}{\sqrt{N}}\left\{(\mathring{R}_1-\hat{R}_1^\ast)\pi_1+(W_1^\ast-\hat{W}_1^\ast)\pi_2+\xi^1\right\}
	\end{align*}} Then following in the manner of the two-period panel, 
	{\small	\begin{align*}
			&\frac{\hat{W}_1^{\ast\prime}\hat{M}^\ast_1\hat{W}_1^\ast}{N} \\
			&= \frac{1}{N}\sum_{i=1}^{N}T_i\Phi^2(R_i\hat{\gamma})\hat{R}_i^{\ast\prime}\hat{R}_i^\ast-\left(\frac{1}{N}\sum_{i=1}^{N}T_i\Phi(R_i\hat{\gamma})\hat{R}_i^{\ast\prime}\hat{R}_i^\ast\right)\left[\frac{1}{N}\sum_{i=1}^{N}T_i\hat{R}_i^{\ast\prime}\hat{R}_i^\ast\right]^{-1}\left(\frac{1}{N}\sum_{i=1}^{N}T_i\Phi(R_i\hat{\gamma})\hat{R}_i^{\ast\prime} \hat{R}_i^\ast\right) \\
			&=\frac{1}{N}\sum_{i=1}^{N}\hat{\Gamma}_{1i}\hat{\Gamma}^\prime_{1i}\overset{p}{\longrightarrow} \mathbb{E}[\Gamma_{1i}\Gamma_{1i}^\prime] = \lambda\cdot \Omega_{\Gamma}
	\end{align*}}
	where $\displaystyle \hat{\Gamma}_{1i} = T_i\Phi(R_i\hat{\gamma})\hat{R}^{\ast\prime}_i-\left(\frac{1}{N}\sum_{i=1}^{N}T_i\Phi(R_i\hat{\gamma})\hat{R}^{\ast\prime}_i\hat{R}^\ast_i\right)\left(\frac{1}{N}\sum_{i=1}^{N}T_i\hat{R}^{\ast\prime}_i\hat{R}^\ast_i\right)^{-1}T_i\hat{R}^{\ast\prime}_i$ and $\displaystyle \Gamma_{1i} = T_i\cdot \Gamma_i = T_i\Phi(R_i\gamma)\mathring{R}_i^\prime-\mathbb{E}[\Phi(R_i\gamma)\mathring{R}_i^\prime\mathring{R}_i]\mathbb{E}[\mathring{R}_i^\prime\mathring{R}_i]^{-1}T_i\mathring{R}_i^\prime$. Therefore, using the above we can write
	{\small \begin{align}\label{if1_pi2}
			\sqrt{N}(\hat{\pi}_2^{2S}- \pi_2)
			& =  \lambda^{-1}\Omega^{-1}_{\Gamma}\left(\frac{\hat{W}_1^{\ast\prime}\hat{M}^\ast_1}{\sqrt{N}}\left\{(\mathring{R}_1-\hat{R}_1^\ast)\pi_1+(W_1^\ast-\hat{W}_1^\ast)\pi_2+\xi^1\right\}\right)+o_p(1)
	\end{align}} 
	Define $\Pi_1(\gamma) = (\Pi_{i1}(\gamma), \ldots, \Pi_{iN}(\gamma))^\prime$ where $\Pi_{i1}^\ast(\gamma) = T_i \cdot \Pi_{i}^\ast(\gamma)$. We can express the terms inside the brackets above as 
	{\small \begin{align}\label{if2_pi2}
			&= \frac{\hat{W}_1^{\ast\prime}\hat{M}^\ast_1}{\sqrt{N}}\left\{(\mathring{R}_1-\hat{R}_1^\ast)\pi_1+(W_1^\ast-\hat{W}_1^\ast+\Pi_1^\ast(\gamma)-\Pi_1^\ast(\gamma))\pi_2+\xi^1\right\} \nonumber \\
			& =  \frac{\hat{W}_1^{\ast\prime}\hat{M}^\ast_1\mathring{R}_1\pi_1}{\sqrt{N}}+\frac{\hat{W}_1^{\ast\prime}\hat{M}^\ast_1\{(W_1^\ast-\Pi_1^\ast(\gamma))\pi_2+\xi^1\}}{\sqrt{N}}+\frac{\hat{W}_1^{\ast\prime}\hat{M}^\ast_1\{\Pi_1^\ast(\gamma)-\hat{W}_1^\ast\}\pi_2}{\sqrt{N}}  \nonumber \\
			& = \sqrt{N}H^\pi_{1N}+\sqrt{N}H^\pi_{2N}+ \sqrt{N}H^\pi_{3N}
	\end{align}}
	where {\small \begin{equation*}
			\begin{split}
				\sqrt{N}H^\pi_{1N} &= \frac{\hat{W}_1^{\ast\prime}\hat{M}^\ast_1\mathring{R}_1\pi_1}{\sqrt{N}} =  \frac{1}{\sqrt{N}}\sum_{i=1}^{N}T_i\hat{\Gamma}_{i}\mathring{R}_i\pi_1 \overset{p}{\longrightarrow} N(0, \Omega_{1\pi})\\
				\Omega_{1\pi}& = \lambda\cdot \mathbb{E}[\Gamma_i\mathring{R}_i\pi_1\pi_1^\prime\mathring{R}_i^\prime \Gamma_i^\prime ] \text{ since } \mathbb{E}[\Gamma_i\mathring{R}_i] = 0
			\end{split}
		\end{equation*}}
	{\small \begin{equation*}
		\begin{split}
				\sqrt{N}H^\pi_{2N} &= \frac{\hat{W}_1^{\ast\prime}\hat{M}^\ast_1\{(W_1^\ast-\Pi_1^\ast(\gamma))\pi_2+\xi^1\}}{\sqrt{N}} = \frac{1}{\sqrt{N}}\sum_{i=1}^{N}T_i\hat{\Gamma}_{i}\{(W_i^\ast-\Pi_i^\ast(\gamma))\pi_2+\xi_{i1}\} \overset{p}{\longrightarrow} N(0, \Omega_{2\pi})\\
				\Omega_{2\pi}& = \lambda\cdot \mathbb{E}[\Gamma_i\{(W_i^\ast-\Pi_i^\ast(\gamma))\pi_2+\xi_{i1}\}\{(W_i^\ast-\Pi_i^\ast(\gamma))\pi_2+\xi_{i1}\}^\prime\Gamma_i^\prime] \text{ since } \mathbb{E}[\Gamma_i\{(W_i^\ast-\Pi_i^\ast(\gamma))\pi_2+\xi_{i1}\}] = 0 \\
				& = \lambda\cdot \{\mathbb{E}[(W_i^\ast-\Pi_i^\ast(\gamma))\pi_2\pi_2^\prime(W_i^{\prime\ast}-\Pi_i^{\prime\ast}(\gamma)) ]+\sigma_1^2\}
		\end{split}
	\end{equation*}} and 
	{\small \begin{equation*}
		\begin{split}
			\sqrt{N}H^\pi_{3N} &= \frac{\hat{W}_1^{\ast\prime}\hat{M}^\ast_1\{\Pi_1^\ast(\gamma)-\hat{W}_1^\ast\}\pi_2}{\sqrt{N}} = \frac{1}{\sqrt{N}}\sum_{i=1}^{N}T_i\hat{\Gamma}_{i}\{\Pi_i^\ast(\gamma)-\hat{W}_i^\ast\}\pi_2 \overset{p}{\longrightarrow} N(0, \Omega_{3\pi})
		\end{split}
	\end{equation*}}
	Using (\ref{if1_pi2}) and (\ref{if2_pi2}) we can write 
	{\small \begin{align}\label{biaspi2} 
			\sqrt{N}(\hat{\pi}_2^{2S}- \pi_2)
			=  \lambda^{-1}\Omega^{-1}_{\Gamma}\bigg(\sqrt{N}H_{1N}^{\pi} +\sqrt{N}H_{2N}^\pi+\sqrt{N}H_{3N}^\pi \bigg)+o_p(1)
	\end{align}} 
	Similarly, 
	{\small \begin{align}\label{biaseta2} 
			\sqrt{N}(\hat{\eta}_2^{2S}- \eta_2)
			=  (1-\lambda)^{-1}\Omega^{-1}_{\Gamma}\bigg(\sqrt{N}H_{1N}^{\eta}+\sqrt{N}H_{2N}^\eta+\sqrt{N}H_{3N}^\eta\bigg)+o_p(1)
	\end{align}} where $\sqrt{N}H_{1N}^\eta\overset{p}{\longrightarrow} N(0, \Omega_{1\eta})$, $\sqrt{N}H_{2N}^\eta\overset{p}{\longrightarrow} N(0, \Omega_{2\eta})$ and $\sqrt{N}H_{3N}^\eta\overset{p}{\longrightarrow} N(0, \Omega_{3\eta})$ are defined analogously.

Therefore, (\ref{biaspi2}) and (\ref{biaseta2}) together give us
	{\small \begin{align}\label{biastheta_pols} 
	\sqrt{N}(\hat{\theta}^{2S}_{POLS}- \theta)
	& =  \Omega^{-1}_{\Gamma}\bigg\{\sqrt{N}\left(\frac{H_{1N}^\pi}{\lambda}-\frac{H_{1N}^\eta}{(1-\lambda)}\right)+\sqrt{N}\left(\frac{H_{2N}^\pi}{\lambda}-\frac{H_{2N}^\eta}{(1-\lambda)}\right) \nonumber \\
	& \ \ \ \ \ \ \ + \sqrt{N}\left(\frac{H_{3N}^\pi}{\lambda}-\frac{H_{3N}^\eta}{(1-\lambda)}\right)\bigg\}+o_p(1) \nonumber \\
	& = \Omega^{-1}_{\Gamma}\frac{1}{\sqrt{N}}\sum_{i=1}^{N}\Gamma_{i}\bigg\{\mathring{R}_i\left(\frac{T_i}{\lambda} \pi_1-\frac{(1-T_i)}{(1-\lambda)}\eta_1\right) +(W_i^\ast-\Pi_i^\ast(\gamma)) \left(\frac{T_i}{\lambda} \pi_2-\frac{(1-T_i)}{(1-\lambda)}\eta_2\right) \nonumber \\
	&+\left(\frac{T_i}{\lambda}\xi_{i1}-\frac{(1-T_i)}{(1-\lambda)}\xi_{i0}\right) +(\Pi_i^\ast(\gamma)-\hat{W}_i^\ast)\cdot \left(\frac{T_i}{\lambda}\pi_2-\frac{(1-T_i)}{(1-\lambda)}\eta_2\right)\bigg\} +o_p(1)\nonumber
	\end{align}} 
\end{proof}

\section{Tables}

\begin{table}[H]\caption{\textbf{Partial Observability Model for BPL Ration Card Ownership}}\label{first_stg}
	\centering
	\begin{threeparttable}
		\footnotesize{
			\begin{tabular}{lcc}
				\toprule\toprule
				& \multicolumn{1}{c}{(1)} & \multicolumn{1}{c}{(2)} \\
				\midrule
				Log operated land &    -0.620* &     -0.080 \\
				&    (0.374) &    (0.443) \\
				Log consumption expenditure &      0.020 &   -0.348** \\
				&    (0.125) &    (0.137) \\
				Any member literate dummy &      0.218 &   -0.350** \\
				&    (0.279) &    (0.146) \\
				Hindu dummy &   0.960*** &            \\
				&    (0.355) &            \\
				Other caste dummy &  -1.131*** &            \\
				&    (0.380) &            \\
				Other backward caste dummy &    -0.373* &            \\
				&    (0.206) &            \\
				Rural dummy &     0.814* &     -0.089 \\
				&    (0.435) &    (0.535) \\
				Any vehicle owned dummy &            &   -0.135** \\
				&            &    (0.061) \\
				Motor vehicle owned dummy &            &  -0.426*** \\
				&            &    (0.100) \\
				Cooler owned dummy &            &  -0.424*** \\
				&            &    (0.132) \\
				TV owned dummy &            &  -0.264*** \\
				&            &    (0.090) \\
				Electric fan owned dummy &            &  -0.417*** \\
				&            &    (0.105) \\
				Refrigerator owned dummy &            &  -0.588*** \\
				&            &    (0.122) \\
				Kaccha house dummy &            &  -0.404*** \\
				&            &    (0.084) \\
				LPG gas dummy &            &   0.471*** \\
				&            &    (0.166) \\
				Tractor/thresher owned dummy &            &  -0.873*** \\
				&            &    (0.169) \\
				Any member in local caste associations &            &   0.590*** \\
				&            &    (0.107) \\
				Panchayat official close to household &            &    0.211** \\
				&            &    (0.083) \\
				Constant &     -0.434 &    3.246** \\
				&    (0.977) &    (1.321) \\
				$\rho$   & \multicolumn{2}{c}{-0.315**} \\
				& \multicolumn{2}{c}{(0.134)} \\
				$N$     & \multicolumn{2}{c}{9380} \\
				Wald test $\rho$=0 & \multicolumn{2}{c}{$\chi^2(1)$ = 4.82, P-value = 0.028} \\
				\bottomrule\bottomrule
			\end{tabular}
			\begin{tablenotes}[flushleft]
				\item \footnotesize Notes: Dependent variable in both equations is a dummy variable which is 1 if the household owns a below poverty line ration card, 0 otherwise. Standard errors in parentheses clustered at the village level. ***, **, and * indicate statistical significance at the 1\%, 5\%, and 10\% levels, respectively.
			\end{tablenotes}
		}
	\end{threeparttable}
\end{table}

\begin{table}[H]\caption{\textbf{Placebo Check}}\label{pll}
	\centering
	\begin{threeparttable}
		\begin{tabular}{lccc}
			\toprule\toprule
			& (1)   & (2)   & (3) \\
			\midrule
			\multicolumn{4}{l}{Dependent variable: Log calories consumer per person per day} \\
			$BPL$   & -0.101*** & -0.039 & -0.039 \\
			& (0.031) & (0.029) & (0.029) \\
			$POST$  & -0.090*** & -0.036** & -0.036** \\
			& (0.020) & (0.014) & (0.014) \\
			$BPL \times POST$ & -0.036 & -0.007 & -0.007 \\
			& (0.034) & (0.030) & (0.030) \\
			&       &       &  \\
			Controls & No & Yes & Yes \\
			Government programs & No & No & Yes \\
			\\
			$N$     & 13656 & 13633 & 13633 \\
			$R^2$    & 0.034 & 0.427 & 0.427 \\
			$F$     & 55.46 & 141.92 & 131.34 \\
			\bottomrule\bottomrule
		\end{tabular}%
		\begin{tablenotes}[flushleft]
			\item \footnotesize Notes: These results are based on DID regressions estimated for the baseline survey. $BPL$  is a dummy variable which is 1 if the household owns a below poverty line ration card, 0 otherwise. $POST$ is a dummy variable which is 1 for households surveyed in 2005 and 0 for households surveyed in 2004. Household controls include log operated land, log consumption expenditure, dummy for a literate member present in the household, caste dummies, number of family members, rural dummy, asset count, dummies for ownership of vehicle, motor car, moterbike, cooler or AC, refrigerator, TV, electric fan, telephone, electricity and finally a dummy for whether the household have a makeshift dwelling or a \textit{kaccha} house. Government programs include participation in the Mahatma Gandhi National Rural Guarantee Scheme (MGNREGS) or benefits received from other welfare programs like health insurance, scholarships, old age pension, maternity scheme, disability scheme, income generation programs other than MGNREGS, assistance from drought/flood compensation, and insurance payouts. Standard errors in parentheses clustered at the village level. ***, **, and * indicate statistical significance at the 1\%, 5\%, and 10\% levels, respectively.
		\end{tablenotes}
	\end{threeparttable}
\end{table}	

\end{document}